\documentclass[preprint,3p]{elsarticle}
\usepackage[toc,page,title,titletoc,header]{appendix}
\usepackage{stmaryrd}
\SetSymbolFont{stmry}{bold}{U}{stmry}{m}{n}
\usepackage{amsmath}
\usepackage{amssymb}
\usepackage{amsthm}
\usepackage{tabularx}
\usepackage{subfigure}
\usepackage{epsfig}
\usepackage{indentfirst}
\usepackage{cases}
\biboptions{sort&compress}
\usepackage{graphicx}
\usepackage{epsfig}
\usepackage{epstopdf}
\usepackage{color}
\usepackage{float}
\usepackage{multirow}
\usepackage{algorithm}
\usepackage{algpseudocode}

\newcommand{\norm}[1]{\lVert#1\rVert}
\renewcommand{\vec}[1]{\mathbf{#1}}

\newcommand{\be}{\begin{equation}}
\newcommand{\ee}{\end{equation}}
\newcommand{\ba}{\begin{array}}
\newcommand{\ea}{\end{array}}
\newcommand{\bea}{\begin{eqnarray}}
\newcommand{\eea}{\end{eqnarray}}
\newcommand{\beas}{\begin{eqnarray*}}
\newcommand{\eeas}{\end{eqnarray*}}

\newtheorem{thm}{Theorem}[section]

\numberwithin{equation}{section}

\begin{document}

\begin{frontmatter}

\title{An Energy-stable Finite Element Method for the Simulation of
Moving Contact Lines in Two-phase Flows}

\author{Quan Zhao}
\ead{matzq@nus.edu.sg}

\author{Weiqing Ren\corref{cor}}
\ead{matrw@nus.edu.sg}
\cortext[cor]{Corresponding author.}
\address{Department of Mathematics, National University of
Singapore, Singapore, 119076}


\begin{abstract}
We consider the dynamics of two-phase fluids, in particular the moving contact line, on a solid substrate. The dynamics are governed by the sharp-interface model consisting of the incompressible Navier-Stokes\slash Stokes equations with the classical interface conditions, the Navier boundary condition for the slip velocity along the wall and a contact line condition which relates the dynamic contact angle of the interface to the contact line velocity. We propose an efficient numerical method for the model. The method combines a finite element method for the Navier-Stokes/Stokes equations on a moving mesh with a parametric finite element method for the dynamics of the fluid interface. The contact line condition is formulated as a time-dependent Robin-type of boundary condition for the interface so it is naturally imposed in the weak form of the contact line model. For the Navier-Stokes equations, the numerical scheme obeys a similar energy law as in the continuum model but up to an error due to the interpolation of numerical solutions on the moving mesh. In contrast, for Stokes flows, the interpolation is not needed so we can prove the global unconditional stability of the numerical method in terms of the energy. Numerical examples are presented to demonstrate the convergence and accuracy of the numerical methods. 
\end{abstract}




\end{frontmatter}
\section{Introduction}

When two immiscible fluids or two phases of one fluid move on a solid substrate,
a moving contact line (MCL) forms at the intersection of the fluid interface and the solid wall.
Modeling and simulation of the MCL have attracted much attention 
in recent years, not only because of many interesting physical phenomena
and associated scientific questions in the problem, but also due to its importance in 
industrial applications, such as ink-jet printing, coating, etc. 
The main difficulty in the problem arises from the well-known stress
singularity at the MCL in classical hydrodynamic models, 
e.g. the Navier-Stokes equations coupled with the conventional no-slip boundary condition
\cite{Huh71, Dussan74}. 
A lot of efforts have been devoted to resolving this difficulty, and
different models have been proposed. These include molecular dynamics models
\cite{Koplik88, Thompson89, Ren07, DeConinck08},
the molecular kinetic theory \citep{Blake69, Blake93}, 
diffuse interface models \citep{Anderson98, Jacqmin00, Pismen02, Qian03, Yue10},
the interface breaking/formation model \citep{Shik97},
and hydrodynamic models 
\cite{Voinov76, Hocking77, Cox86,  Eggers04a, Ren10, Ren11d, Ren15, ZhangRen2019, Sibley15}.
We refer to the review articles \cite{Dussan79, deGennes85, Kistler93, Pomeau02, Bonn09}, 
the collected volume \cite{Velarde11} 
and the monographs \cite{deGennes03, Starov07} 
for details of these different models and discussions of the MCL problem.


In addition to the work on modelling MCLs, there also exists a large body 
of numerical work in the literature, e.g., 
\cite{Afkhami09,Renardy01, Dupont2010, Li10, Ren11,Spelt05, Zahedi09, Xu14,Xu16, 
Gao2014,Bao2012finite, Ding2008, Carlson2009, Zhang16,Huang04, Muradoglu10, Zhang14}. 
The readers are referred to the review article \cite{Sui14} for detailed discussions. 
These methods use different methods to represent the fluid interface and/or
different contact line conditions as well as their numerical implementations. 
For example, in Refs. \cite{Afkhami09,Renardy01, Dupont2010}, the volume of fluid method was used
to deal with the moving interface and the contact angle condition was imposed 
on the gradient of the volume fraction function at the contact line. Traditional interface-capturing methods have been extended to systems with MCLs, 
including the level set method~\cite{Li10, Ren11,Spelt05, Zahedi09, Xu14,Xu16} and the diffuse interface approach~\cite{Gao2014, Bao2012finite, Ding2008, Carlson2009}. Li {\it et al.} proposed an augmented immersed interface method and employed a prescribed
profile for the slip velocity near the MCL \cite{Li10}. 
Spelt proposed a macroscale approach to simulate MCLs with hysteresis where
the contact line only moves when the dynamic contact angle is not 
within a prescribed region \cite{Spelt05}. 
Bao {\it et al.} proposed a finite element method for the coupled Cahn-Hillard 
and Navier-Stokes equations with generalized Navier boundary condition 
for the MCLs \cite{Bao2012finite}. The front tracking method, in which the interface was represented by a number of markers, 
can be found in Refs. \cite{Huang04,Muradoglu10, Zhang14}, 
and the contact line position is updated according to either the fluid velocity at the contact line
 or the contact angle.

In this work, we will restrict ourselves to the contact line model proposed 
by Ren {\it et al.}~\cite{Ren07,Ren10,Ren11d}. This is a sharp interface model and was
developed based on molecular dynamics simulations and the consideration 
of thermodynamics laws. It consists of the incompressible Navier-Stokes equations 
with the classical interface conditions, the Navier slip condition at the wall 
and a contact line condition. The contact line condition can be viewed as a force balance,
in which the friction force at the contact line is balanced by the stress resulted 
from the deviation of the dynamic contact angle from its equilibrium value. 
The latter is usually referred to as the unbalanced Young stress. 
In the earlier work \cite{Ren11, Xu14}, the contact line condition was unified with 
the Navier slip condition by applying a singular force at the contact line.
The resulting condition was then applied to the Navier-Stokes equations 
to determine the velocity field including the slip velocity along the whole solid wall. 
This approach is similar 
to the continuum force method for the simulation of multi-phase flows 
where the interface conditions
are imposed by applying singular forces along the interface in the momentum equation.

In the current work, we propose a finite element method (FEM), 
based on the earlier work of Barrett {\it et al.} \cite{Barrett15stable}. 
The earlier work dealt with multi-phase flows with closed interfaces. 
Here we extend it systems to with moving contact lines. 
In the numerical method, an efficient finite element discretization 
for the Navier-Stokes\slash Stokes equations is coupled with 
a parametric finite element approximation for the fluid interface. 
The contact line condition is naturally imposed by using the weak form of 
the governing equations.

The contact line model obeys an energy law: The total energy, including the kinetic
energy and the interface energies, is dissipated due to the viscous stress 
in the bulk of the fluids, the friction force on the wall and the contact line friction.
So it is desirable that the numerical method has a similar property.
Indeed, for the FEM we can establish a similar energy law but up to interpolation errors. 
We use a moving mesh approach so that the mesh remain fitted to the evolving 
fluid interface. This requires the interpolation 
of the velocity and density fields which were solved on the mesh 
at the previous time step to the new mesh at the current time step. 
The induced interpolation error pollutes the numerical solution; 
as a result, we can only establish an energy bound locally at each time step. 
In contrast, for Stokes equations, the interpolation of the solutions is not needed,  
and the corresponding FEM enjoys a global energy bound.

The rest of the paper is organized as follows. In section \ref{sec:model}, we
review the contact line model, including the governing equations 
and boundary/interface conditions, and then propose a weak formulation for the model. 
In section \ref{sec:num}, we propose the numerical method 
based on the weak form of the model, prove the well-posedness and an energy bound for 
the numerical scheme and a moving mesh approach for the generation of the 
fitted mesh. Subsequently, in section \ref{sec:results}
we report some numerical results to demonstrate 
the convergence and accuracy of the numerical method. 
In section \ref{sec:stokes}, we consider the case when the flow is modelled 
by the time-independent Stokes equations. We present the corresponding 
numerical method and demonstrate its convergence and accuracy using numerical examples.
Finally, we draw the conclusion in section \ref{sec:con}.

\section{The contact line model and its weak formulation}
\label{sec:model}
In this section, we first review the moving contact line model proposed by Ren {\it et al}~\cite{Ren10} 
and introduce the dimensionless governing equations with dimensionless boundary and interface 
conditions. We then present a weak formulation for the dimensionless model. 

\subsection{Governing equations}
\begin{figure}[!htp]
\centering
\includegraphics[width=0.80\textwidth]{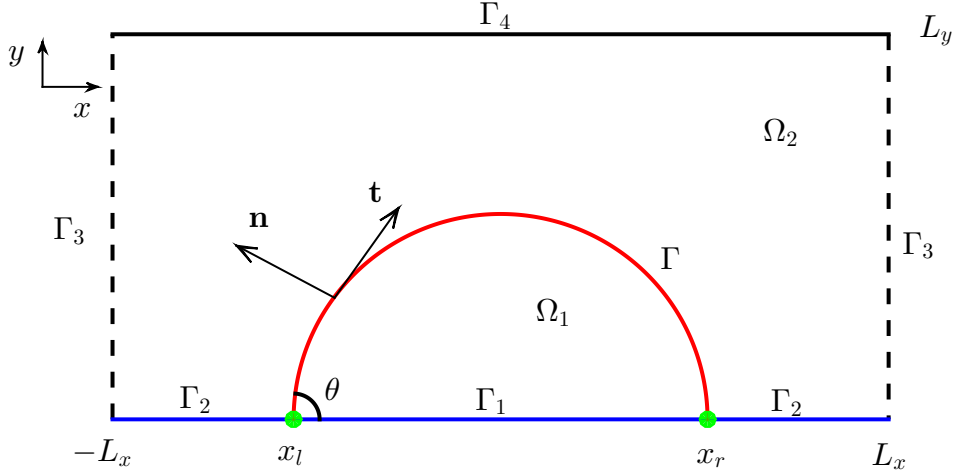}
\caption{A schematic illustration of the moving contact lines (green points labeled as $x_l$ and $x_r$) in two-phase flows in a bounded domain $\Omega=\Omega_1\cup\Omega_2=[-L_x,~L_x]\times[0,~L_y]$, where the red solid line and blue solid line represent the fluid interface $\Gamma$ and the rigid solid substrate $\Gamma_1\cup\Gamma_2$, respectively. }
\label{fig:model}
\end{figure}

Without loss of generality, we consider the dynamics of a liquid droplet
on a stationary solid substrate in the 2d space, as shown in Fig.~\ref{fig:model}. 
We use Cartesian coordinates, where the substrate is on the $x$ axis. 
The physical domain $\Omega$ consists of two regions: one is occupied by the droplet
and denoted by $\Omega_1$, the other is occupied by the fluid outside the droplet 
and denoted by $\Omega_2$.  

Let $\rho_i\ (i=1,2)$ denote the density of the fluids, 
$\vec u(\vec x,~t):\;\Omega\times[0,T]\rightarrow\mathbb{R}^2$ be the fluid velocity, 
and $p(\vec x,~t):\;\Omega\times[0,T]\rightarrow\mathbb{R}$ be the pressure.
The dynamics of the system is governed by the standard incompressible Navier-Stokes 
equations in $\Omega_i$ ($i=1,2$),
\begin{subequations}
\begin{numcases}{}
\label{eqn:fluidynamic1}
\rho_i\,(\partial_t\vec u + \vec u\cdot\nabla\vec u) = -\nabla p + \nabla\cdot\tau_d,\\[0.5em]
 \nabla\cdot\vec u = 0,
 \label{fluidynamic2}
\end{numcases}
\end{subequations}
where $\tau_d = 2\eta_i D(\vec u)$ is the viscous stress with 
$D(\vec u)=\frac{1}{2}(\nabla\vec u + (\nabla\vec u)^T)$, 
and $\eta_i\ (i=1,2)$ are the viscosities of the fluids.


On the fluid interface $\Gamma(t)$, we have the following conditions hold
 \begin{subequations}
 \begin{align}
 \label{eqn:fc1}
  \bigl[\vec u\bigr]^2_1 = 0,\\
   \label{eqn:fc2}
 \bigl[p\mathbf{I}-\tau_d\bigr]_1^2\cdot\vec n = \gamma\kappa\vec n,\\
 \dot{\vec x}_\Gamma =\vec u |_{\vec x_\Gamma},
 \label{eqn:fc3}
 \end{align}
 \end{subequations}
where $\bigl[\cdot\bigr]_1^2$ denotes the jump from fluid 1 to fluid 2, 
$\mathbf{I}\in\mathbb{R}^{2\times2}$ is the identity matrix, 
$\gamma$ is the surface tension of the fluid interface, $\vec n$ and $\kappa$ 
are the unit normal vector and curvature of the fluid interface $\Gamma$ respectively,
and $\dot{\vec x}_\Gamma$ denotes the velocity of the fluid interface.
Eq.~\eqref{eqn:fc1} states that the fluid velocity is continuous across the interface, 
Eq.~\eqref{eqn:fc2} is the balance of the normal stress jump of the fluids
and the capillary force, and 
Eq. \eqref{eqn:fc3} is the kinematic condition for the interface.


At the lower solid wall $\Gamma_1\cup\Gamma_2$, the fluid velocity satisfies 
 the no-penetration condition and the Navier boundary condition
\begin{subequations}  
\begin{align} \label{eq:uwn}
\vec u\cdot\vec n_w = 0, \\
\label{eqn:slipbd1}
\vec t_w\cdot\tau_d\cdot\vec n_w = -\beta_iu_s, 
\end{align}
\end{subequations}
where $\vec n_w=(0,-1)^T$ and $\vec t_w=(1,0)^T$ are unit normal and tangent vectors 
of the wall, respectively; $\beta_i\ (i=1,2)$ are the friction coefficients 
of the fluids at the solid wall, and  $u_s=\vec u\cdot\vec t_w$ is the slip velocity
of the fluids. The dynamic contact angles $\theta_d^l$ and $\theta_d^r$ 
that the fluid interface forms with the solid wall satisfy
\begin{subequations}
\begin{align}
\label{eqn:slipbd2}
& \gamma\left(\cos(\theta_d^l)-\cos\theta_Y\right)=\beta^{*}\dot{x}_l, \\
\label{eqn:slipbd3}
&\gamma\left(\cos(\theta_d^r) - \cos\theta_Y\right) = - \beta^* \dot{x}_r, 
\end{align}
\end{subequations}
where $\beta^{*}$ is the friction coefficient of the fluid interface at the solid wall,
$\dot{x}_l$ and $\dot{x}_r$ are the velocities of the contact points, 
and $\theta_Y$ is the equilibrium contact angle satisfying the Young's relation
\begin{equation}
\gamma\cos\theta_Y=\gamma_2-\gamma_1,
\end{equation}
where $\gamma_1$ and $\gamma_2$ are the surface tension coefficient
at the interface  $\Gamma_1$ and $\Gamma_2$, respectively. We note that 
since the fluid interface evolves with the fluid velocity according to
Eq.~\eqref{eqn:fc3}, we have $\dot{x}_{l,r} = u_s\big|_{x=x_{l,r}}$.
Finally, we use the no-slip condition at the upper wall $\Gamma_4$ and periodic 
conditions at $\Gamma_3$. 
%

The total energy of the system is given by
 \begin{equation}
 E(t)
= \sum_{i=1,2}\int_{\Omega_i(t)}
\frac{1}{2}\rho_i|\vec u|^2\;d\mathcal{L}^2 +(\gamma_1-\gamma_2)|\Gamma_1(t)| 
+ \gamma|\Gamma(t)|,
 \end{equation}
where $|\Gamma_1(t)|$ and $|\Gamma(t)|$ denote the arc length of the line segment 
$\Gamma_1(t)$ and the curve $\Gamma(t)$, respectively. The three terms
represent the kinetic energy of the fluids, 
the interracial energy at the solid wall and the interfacial energy of the fluid interface,
respectively. The dynamical system obeys the following energy dissipation law 
\cite{Ren07, Ren11, Ren11d}:
\begin{equation}
\frac{{\rm d}}{{\rm d}t}E(t) = -\sum_{i=1,2}\int_{\Omega_i}\eta_i|\nabla\vec u|^2\;d\mathcal{L}^2-\sum_{i=1,2}\int_{\Gamma_i}\beta_i|u_s|^2\;ds -\beta^*
\left(\dot{x}_l^2 + \dot{x}_r^2 \right) \le 0.
\label{eqn:energydissipation2d}
\end{equation}

\subsection{Dimensionless equations}
Next, we write the governing equations and boundary/interface conditions 
in their dimensionless form. We rescale the physical quantities as 
\begin{equation*}
\hat{\rho}_i = \frac{\rho_i}{\rho_2},\quad \hat{\eta_i} = \frac{\eta_i}{\eta_2}, 
\quad\hat{\beta_i} = \frac{\beta_i}{\beta_2}, \quad \hat{\beta^*} = \frac{\beta^*}{\eta_2},
\quad\hat{\gamma}_i = \frac{\gamma_i}{\gamma},
\end{equation*}
\begin{equation*}
\hat{\vec x} = \frac{\vec x}{L},\quad \hat{\vec u} = \frac{\vec u}{U},\quad\hat{t} 
= \frac{Ut}{L},\quad \hat{p} = \frac{p}{\rho_2 U^2},\quad \hat{\kappa} = L\kappa, 
\end{equation*}
where $L$ and $U$ are the characteristic length and velocity, respectively.
We define the Reynolds number $Re$, the Capillary number $Ca$, the slip length $l_s$, 
and the Weber number $We$ as follows,
\begin{equation*}
Re = \frac{\rho_2UL}{\eta_2},\quad Ca = \frac{\eta_2U}{\gamma}, 
\quad l_s = \frac{\eta_2}{\beta_2L},\quad We = Re\cdot Ca.
\end{equation*}
Then the governing equations in $\Omega_i\ (i=1,2)$ 
can be rewritten as (dropping the hats):
\begin{subequations} \label{eq:model12}
\begin{numcases}{}
\label{eqn:model1}
\rho_i\,(\partial_t\vec u + \vec u\cdot\nabla\vec u) +\nabla\cdot\sigma =0,\\
\label{eqn:model2}
\nabla\cdot\vec u = 0,
\end{numcases}
\end{subequations}
where $\sigma = p\mathbf{I} - \frac{1}{Re}\tau_d$.
%
%
The above governing equations are coupled with the following boundary/interface conditions:
\begin{itemize}
\item [(i)] The interface conditions on $\Gamma(t)$:
\begin{subequations} \label{eq:bd1}
\begin{align}
  \bigl[\vec u\bigr]^2_1 = 0,\\
We\,\bigl[\sigma\bigr]_1^2\cdot\vec n = \kappa\,\vec n,\\
\kappa = (\partial_{ss}\vec X)\cdot\vec n, \label{eq:cv}\\
 \dot{\vec x}_\Gamma = \vec u|_{\vec x_\Gamma}.
  \label{eq:k}
\end{align}
\end{subequations}
where $s$ is the arc-length parameter of the fluid interface.

\item [(ii)] The boundary conditions on $\Gamma_1(t)\cup\Gamma_2(t)$:
\begin{subequations}  \label{eq:bd2ab}
\begin{align}
\label{eqn:bd2}
\vec u\cdot\vec n_w=0,\\
l_s\,\vec t_w\cdot\tau_d\cdot\vec n_w =  -\beta_i u_s.
\end{align}
\end{subequations}
\item [(iii)] The condition for the dynamic contact angles:
\begin{subequations} \label{eqn:bd3}
\begin{align}
& \frac{1}{Ca}(\cos\theta_d^l - \cos\theta_Y)=\beta^*\dot{x}_l(t), \\
& \frac{1}{Ca}(\cos\theta_d^r - \cos\theta_Y)=-\beta^*\dot{x}_r(t).
\end{align}
\end{subequations}
%
\item [(iv)] Periodic boundary conditions on $\Gamma_3$:
\begin{subequations} \label{eq:bd5ab}
\begin{align}
\label{eqn:bd5}
\vec u(-L_x,y,t) = \vec u(L_x,y,t),\\
\sigma(-L_x,y,t) = \sigma(L_x,y,t).
\end{align}
\end{subequations}
\item[(v)] The no-slip condition on the upper wall $\Gamma_4$:
\begin{equation}
\label{eqn:bd6}
\vec u = \vec 0.
\end{equation}
\end{itemize}

In terms of the dimensionless variables, the total energy 
(rescaled by $\rho_2U^2L^2$) of the system  becomes
\begin{equation}
\label{eqn:dimenenergy}
E(t)
=\sum_{i=1,2}\int_{\Omega_i(t)}\frac{1}{2}\rho_i|\vec u|^2\;
d\mathcal{L}^2 -\frac{\cos\theta_Y}{We}|\Gamma_1(t)| + \frac{1}{We}|\Gamma(t)|,
\end{equation}
and the system  obeys the energy dissipation law
\begin{equation}
\frac{{\rm d}}{{\rm d}t}E(t) = -\sum_{i=1,2}\frac{1}{Re}\int_{\Omega_i}
\eta_i|\nabla\vec u|^2\;d\mathcal{L}^2-\sum_{i=1,2}\frac{1}{Re\,l_s}
\int_{\Gamma_i}\beta_i|u_s|^2\;ds -\frac{\beta^{*}}{Re}
\left(\dot{x}_l^2 + \dot{x}_r^2\right)\le 0.
\label{eqn:dimensionenergydissipation2d}
\end{equation}

\subsection{Weak formulation} \label{sec:v}

In order to propose the weak formulation for equations \eqref{eq:model12} - 
\eqref{eqn:bd6}, we define the following function space for the fluid velocity,
\begin{equation}
\mathbb{U}:=\left\{\boldsymbol{\omega}\in \left[H^1(\Omega)\right]^2:\;
\boldsymbol{\omega}\cdot\vec n_w =0\;{\rm on}\;\Gamma_1\cup\Gamma_2,\;
\boldsymbol{\omega}=\vec 0\;{\rm on}\;\Gamma_4,\;{\rm and}\;
\boldsymbol{\omega}(-L_x,~y) = \boldsymbol{\omega}(L_x,~y)\right\},
\label{eqn:USpace}
\end{equation}
and the following function spaces for the pressure,
\begin{equation}  \label{eq:P}
\mathbb{P}:=\left\{\varphi\in L^2(\Omega)\right\},
\qquad\hat{\mathbb{P}}:=\left\{\varphi\in\mathbb{P}:\;
\int_\Omega \varphi d\mathcal{L}^2=0\right\}.
\end{equation}
We parameterize the fluid interface as $\vec X(\alpha, t)=(X(\alpha, t), Y(\alpha, t))$,
where $\alpha\in I=[0,1]$, and $\alpha=0,\ 1$ correspond to the left and right 
contact point, respectively. We define the following function space with respect to 
the interface, 
\begin{equation}
L^2(I)=\left\{u: I\rightarrow \mathbb{R}, \;\text{and} 
\int_I |u(\alpha)|^2 |\partial_\alpha\vec{X}|\, d\alpha <+\infty \right\},
\end{equation}
equipped with the inner product
\be
\big(u,v\big)_{\Gamma} =
\int_{I}u(\alpha) v(\alpha)|\partial_\alpha\vec{X}|\,d\alpha,\quad \forall\;u,v\in L^2(I).
\ee


We take the inner product of Eq. \eqref{eqn:model1} with $\boldsymbol{\omega}$,
for $\forall\boldsymbol{\omega}\in\mathbb{U}$.
Using the boundary/interface conditions in \eqref{eq:bd1}, \eqref{eq:bd2ab}, 
\eqref{eq:bd5ab} and \eqref{eqn:bd6}, as well as $\nabla\cdot\vec u=0$,
we have \cite{Barrett15stable,Barrett15stable2}
\begin{equation}
\Bigl(\rho\,[\partial_t\vec u + (\vec u\cdot\nabla)\vec u], \boldsymbol{\omega}\Bigr)
= \frac{1}{2}\,\left[\frac{\rm{d}}{\rm{d}t}\Bigl(\rho\,\vec u, \boldsymbol{\omega}\Bigr)
+\Bigl(\rho\,\partial_t\vec u, \boldsymbol{\omega}\Bigr)\right] + \frac{1}{2}\,
\Bigl(\rho, [(\vec u\cdot\nabla)\vec u]\cdot\boldsymbol{\omega}
-[(\vec u\cdot\nabla)\boldsymbol{\omega}]\cdot\vec u\Bigl),
\label{eqn:immediate1}
\end{equation}
where  $\rho = \rho_1\chi_{_{\Omega_1}}+\rho_2\chi_{_{\Omega_2}}$,
$\chi$ is the characteristic function, and
$(\cdot,\cdot)$  denotes the $L^2$ 
inner product on $\Omega_1\cup \Omega_2$,
\begin{equation*}
(\vec u, \vec v)=\sum_{i=1,2}\int_{\Omega_i} \vec u\cdot\vec v  d\mathcal{L}^2.
\end{equation*}
For the viscous term, take the inner product with $\boldsymbol{\omega}\in\mathbb{U}$. We use $\sigma=p\mathbf{I} - \frac{1}{Re}\tau_d$, and 
apply integration by parts, which yields
\begin{align}
\Bigl(\nabla\cdot\sigma, \boldsymbol{\omega}\Bigr)
&=-\Bigl(p, ~\nabla\cdot\boldsymbol{\omega}\Bigr) + 
 \frac{2}{Re}\Bigl(\eta D(\vec u), ~D(\boldsymbol{\omega})\Bigr)
 - \Bigl([\sigma]_1^2\cdot\vec n, ~\boldsymbol{\omega}\Bigr)_\Gamma 
+ \Bigl(\sigma\cdot\vec n_w,~\boldsymbol{\omega}\Bigr)_{\Gamma_1\cup\Gamma_2} \nonumber\\
&= -\Bigl( p,~\nabla\cdot\boldsymbol{\omega}\Bigr) + \frac{2}{Re}
\Bigl(\eta D(\vec u), D(\boldsymbol{\omega})\Bigr) -\frac{1}{We}
\Bigl(\kappa\,\vec n, \boldsymbol{\omega}\Bigr)_\Gamma 
- \frac{1}{Re}\Bigl(\tau_d\cdot\vec n_w, \boldsymbol{\omega}\Bigr)_{\Gamma_1\cup\Gamma_2}
 \nonumber\\
&=-\Bigl( p, \nabla\cdot\boldsymbol{\omega}\Bigr) 
 + \frac{2}{Re}\Bigl(\eta D(\vec u), D(\boldsymbol{\omega})\Bigr) 
 -\frac{1}{We}\Bigl(\kappa\,\vec n, \boldsymbol{\omega}\Bigr)_\Gamma 
 + \frac{1}{Re\,l_s}\Bigl(\beta\,u_s, \omega_s)_{\Gamma_1\cup\Gamma_2},
\label{eqn:immediate2}
\end{align} 
where $\eta = \eta_1\chi_{_{\Omega_1}} + \eta_2\chi_{_{\Omega_2}}$, 
$\beta = \beta_1\chi_{_{\Gamma_1}} + \beta_2\chi_{_{\Gamma_2}}$, 
$\omega_s=\boldsymbol{\omega}\cdot\vec t_w$, and we have used the boundary and interface
conditions and the fact that 
$\boldsymbol{\omega} = (\boldsymbol{\omega}\cdot\vec t_w)\,\vec t_w =\omega_s\,\vec t_w$ 
on $\Gamma_1\cup\Gamma_2$.

Equation \eqref{eq:cv} for the curvature can be rewritten as 
$\kappa\,\vec n =\partial_{ss}\vec X$. Multiplying this equation 
by a test function $\vec g=(g_1, g_2)\in H^1(I)\times H_0^1(I)$ 
then integrating over $\Gamma(t)$ yields
\begin{align}
0 &=\Bigl(\kappa, \vec n\cdot\vec g\Bigr)_\Gamma
+\Bigl(\partial_s\vec X, \partial_s\vec g\Bigr)_\Gamma 
- (\partial_s\vec X\cdot\vec g)\Big|_{\alpha=0}^{\alpha=1}\nonumber\\
&=\Bigl(\kappa, \vec n\cdot\vec g\Bigr)_\Gamma
 +\Bigl(\partial_s\vec X, \partial_s\vec g\Bigr)_\Gamma 
-(g_1 \partial_sX)\Big|_{\alpha=0}^{\alpha=1}\nonumber\\
&=\Bigl(\kappa, \vec n\cdot\vec g\Bigr)_\Gamma
+\Bigl(\partial_s\vec X, \partial_s\vec g\Bigr)_\Gamma 
+ \beta^*\,Ca\left[\dot{x}_l g_1(0) + \dot{x}_r g_1(1)\right] 
 - \cos\theta_Y \left[g_1(1)-g_1(0)\right],
\label{eqn:curvaturefor}
\end{align}
where we have used fact that $g_2(0)=g_2(1)=0$ in the second equality,
and $\partial_s X|_{\alpha=0}=\cos\theta_d^l,\,\partial_s X|_{\alpha=1}=\cos\theta_d^r$
and the contact angle condition \eqref{eqn:bd3} in the last equality.

From these results, we obtain the weak formulation for the dynamic system  
Eqs.~\eqref{eq:model12}-\eqref{eqn:bd6} as follows: 
Given the initial fluid velocity $\vec u_0$ and interface $\vec X_0(\alpha)$, 
find the fluid velocity $\vec u(\cdot,~t)\in \mathbb{U}$, the pressure 
$p(\cdot,~t)\in \hat{\mathbb{P}}$, the fluid interface 
$\Gamma(t):=\vec X(\cdot,~t)\in\ H^1(I)\times H_0^1(I)$, 
and the curvature  $\kappa(\cdot,~t)\in L^2(I)$ such that 
\begin{subequations}
\begin{align}
\label{eqn:weak1}
&\frac{1}{2}\,\Bigl[\frac{{\rm d}}{{\rm d}t}\Bigl(\rho\,\vec u,~\boldsymbol{\omega}
\Bigr)+\Bigl(\rho\,\partial_t\vec u,~\boldsymbol{\omega}\Bigr)
+ \Bigl(\rho\,(\vec u\cdot\nabla)\vec u,~\boldsymbol{\omega}\Bigr)
-\Bigl(\rho\,(\vec u\cdot\nabla)\boldsymbol{\omega},~\vec u\Bigr)\Bigr] 
+\frac{2}{Re}\,\Bigl(\eta D(\vec u),~D(\boldsymbol{\omega})\Bigr) \nonumber\\
&\qquad\qquad-\Bigl(p,~\nabla\cdot\boldsymbol{\omega}\Bigr) -\frac{1}{We}
\Bigl(\kappa\,\vec n,~\boldsymbol{\omega}\Bigr)_\Gamma +\,\frac{1}{Re\,l_s}
\Bigl(\beta\, u_s,~\omega_s\Bigr)_{\Gamma_1\cup\Gamma_2}=0,
\qquad\forall\boldsymbol{\omega}\in \mathbb{U},\\[0.7em]
\label{eqn:weak2}
&\qquad\qquad\qquad\qquad\qquad\qquad\qquad\Bigl(\nabla\cdot\vec u,~q\Bigr)=0,
\qquad\forall q\in \hat{\mathbb{P}}, \\[0.7em]
\label{eqn:weak3}
&\qquad\qquad\qquad\Bigl(\partial_t\vec X\cdot\vec n,~\psi\Bigr)_{\Gamma} 
- \Bigl(\vec u\cdot\vec n,~\psi\Bigr)_{\Gamma}=0,\qquad\forall 
 \psi\in L^2(I), \\[0.7em]
\label{eqn:weak4}
&\Bigl(\kappa\,\vec n,~\boldsymbol{g}\Bigr)_\Gamma+\Bigl
(\partial_s\vec X,~\partial_s\boldsymbol{g}\Bigr)_\Gamma+ 
\beta^*Ca\Bigl[\dot{x}_l g_1(0) + \dot{x}_r g_1(1)\Bigr] 
-\cos\theta_Y [g_1(1) - g_1(0)]=0, \nonumber \\
& \qquad\qquad\qquad\qquad \hspace{7cm}\forall\boldsymbol{g} \in H^1(I)\times H_0^1(I).
\end{align}
\end{subequations}
Eq. \eqref{eqn:weak1} is a direct result from Eq.~\eqref{eqn:immediate1} 
and Eq.~\eqref{eqn:immediate2}. Eq. \eqref{eqn:weak2} is from the incompressibility
condition. Eq.~\eqref{eqn:weak3} is obtained from the kinematic condition
\eqref{eq:k}, after rewriting it as $\partial_t\vec X\cdot\vec n = \vec u\cdot \vec n$
with $Y(0)=Y(1)=0$. Eq.~\eqref{eqn:weak4} is obtained from Eq.~\eqref{eqn:curvaturefor}. 

The system \eqref{eqn:weak1} - \eqref{eqn:weak4} is an extension 
of the weak formulation introduced in Ref. \cite{Barrett15stable} for 
two-phase flows. Here we have extended it two-phase flows
with moving contact lines. One can prove the energy dissipation 
and mass/area conservation properties within the weak formulation in a similar manner
as did in Ref. \cite{Barrett15stable}.

\section{The numerical method}
\label{sec:num}
Next, we present a finite element method (FEM) based on the weak formulation 
\eqref{eqn:weak1}-\eqref{eqn:weak4} and show the well-posedness and stability 
for the discretized system. Moreover, we propose a moving mesh approach 
for the construction of the mesh such that the fluid interface 
remains fitted to the mesh at each time step.

\subsection{The finite element method}   \label{subsec:fem}

We partition the  time domain $[0,T]$ as $0=t_0<t_1<t_2<\cdots<t_M=T$ with 
the time steps $\tau_m=t_{m+1}-t_m\;(m=0,\cdots,M-1)$ and the 
reference domain $I=[0,1]$ for the fluid interface  as
$I=\bigcup_{j=1}^{J_{_\Gamma}}I_j$, where $I_j=[\alpha_{j-1},\alpha_{j}]$ with
$\alpha_{j}=jh$ and $h=1/J_{_\Gamma}$. We use the following finite-dimensional 
spaces to approximate $H^1(I)$ and $H^1_0(I)$, respectively,
\begin{subequations}
\begin{align}
\label{eqn:FEMspace1}
V^h:&=\left\{u\in C(I):\;u\mid_{I_{j}}\in \mathcal{P}_1(I_j),
\quad \forall \, j=1,2,\ldots, J_{_\Gamma}\right\} \\ 
V_0^h: & = \left\{ u\in V^h: u(0)=u(1)=0\right\},
\end{align}
\end{subequations}
where $\mathcal{P}_1$ denotes the space of polynomials with degrees at most 1.

Let $\Gamma^m:=\vec X^m(\cdot)\in V^h\times V_0^h$ be the numerical approximation 
to the fluid interface $\Gamma$ at the time $t=t_m$. 
For piecewise continuous functions $u$ and $v$ defined on the interval $I$ 
with possible jumps at the nodes 
$\{\alpha_j\}_{j=1}^{J_{_\Gamma}-1}$, we approximate the inner product $(u,v)_{\Gamma(t_m)}$
by either the Simpson rule $\big(u,~v\big)_{\Gamma^m}$ or the Trapezoidal rule $\big(u, v\big)_{\Gamma^m}^h$ (the mass-lumped norm) as
\begin{align}
\label{eqn:simpson}
&\big(u, v\big)_{\Gamma^m}:=\frac{1}{6}\sum_{j=1}^{J_\Gamma}\Big|\vec{X}^m(\alpha_{j})-
\vec{X}^m(\alpha_{j-1})\Big|\Big[\big(u\cdot v\big)(\alpha_{j-1}^+)
+4\big(u\cdot v\big)(\alpha_{j-\frac{1}{2}})+\big(u\cdot v\big)(\alpha_j^-)\Big],\\
\label{eqn:massnorm}
&\big(u, v\big)_{\Gamma^m}^h:=\frac{1}{2}\sum_{j=1}^{J_\Gamma}\Big|\vec{X}^m(\alpha_{j})-
\vec{X}^m(\alpha_{j-1})\Big|\Big[\big(u\cdot v\big)(\alpha_{j-1}^+)
+\big(u\cdot v\big)(\alpha_j^-)\Big],
\end{align}
where $u(\alpha_j^\pm)$ are the one-sided limits of $u$ at $\alpha_j$ and $\alpha_{j-\frac{1}{2}} = \frac{1}{2}(\alpha_{j-1}+\alpha_j)$.
Let $\vec{n}^m$ and $\kappa^m$ be the numerical approximations to the normal vector 
and the curvature of $\Gamma(t_m)$, respectively. 
On each interval $I_j$, the normal vector $\vec{n}^m$ is a constant vector and is 
computed as
\begin{equation}
\vec n_j^m:= \vec n^m\Big|_{I_j} = [\partial_s\vec X^m]^{\perp}\Big|_{I_j}
=\frac{\left[\vec X^m(\alpha_{j}) - \vec X^m(\alpha_{j-1})\right]^\perp}
{\left|\vec X^m(\alpha_j) - \vec X^m(\alpha_{j-1})\right|},\quad 1\leq j\leq J_{_\Gamma},
 \end{equation}
where $(\cdot)^\perp$ denotes the counterclockwise rotation by $\frac{\pi}{2}$. 
In the following, we shall assume $\forall0\leq m\leq M$,
\begin{subequations}
\label{eqn:assumption}
\begin{align}
&\Gamma^m\;\text{has no self-intersections},\\
&n^{m,1}_1\ne 0,\quad n^{m,1}_{J_{_\Gamma}}\ne 0,\\
&\left|\partial_\alpha\vec X^m\right|>0,
\end{align}
\end{subequations}
where $\vec n^m_j = \bigl(n_j^{m,1}, n_j^{m,2}\bigr)$. 
These conditions imply that (1) the first and last line segments of
$\Gamma^m$ are not parallel to the $x$-axis; 
(2) the mesh points on $\{\Gamma^m\}_{m=1}^M$ do not merge. 

Let  $\mathcal{T}^m:=\bigcup_{j=1}^N\bar{o}_j^m$
be a triangulation of $\Omega$ at the time step $t=t_m$. 
The mesh contains $J_{\Omega}$ vertices 
denoted by $\left\{\vec q_k^m\right\}_{k=1}^{J_{\Omega}}$. 
We use a fitted mesh such that the interface $\Gamma^m$ is fitted to 
the triangular mesh $\mathcal{T}^m$. Specifically, the line segments of $\Gamma^m$ 
are edges of triangles from the mesh, i.e., 
$\Gamma^m\subset\bigcup_{j=1}^{N}\partial o_j^m$.
We define the following finite element spaces over $\mathcal{T}^m$,
\begin{subequations}
\begin{align}
S_k^m:& =\left\{\varphi_h\in C(\bar{\Omega}):\varphi_h|_{o_j^m}\in 
\mathcal{P}_k(o_j^m),\; j=1,\cdots,N\right\}, \\
S_0^m:& = \{\varphi_h\in L^2(\Omega):\varphi_h|_{o_j^m}\in \mathcal{P}_0(o_j^m),\;
 j=1,\cdots,N\},
\end{align}
\end{subequations}
where $k\in \mathbb{N}^{+}$, and $\mathcal{P}_k(o_j^m)$ denotes the space of 
polynomials of degree k on $o_j^m$. 

The interface $\Gamma^m$ divides the domain $\Omega$ into $\Omega_1^m$ 
and $\Omega_2^m$. Correspondingly, the mesh $\mathcal{T}^m$ is divided into 
$\mathcal{T}_1^m$ and $\mathcal{T}_2^m$, which consist of triangles in $\Omega_1^m$
and $\Omega_2^m$, respectively.
%
%
Based on the spatial discetization, we define the friction coefficient 
$\beta^m$ and the viscosity $\eta^m\in S_0^m$ as
%
\begin{equation}
\beta^m = \beta_1\chi_{_{\Gamma_1^m}} +\beta_2\chi_{_{\Gamma_2^m}},
\quad \eta^m = \eta_1\chi_{_{\Omega_1^m}} +\eta_2\chi_{_{\Omega_2^m}}.
\end{equation}
Moreover, we define the density $\rho^m\in S_1^m$ such that 
at the vertices $\{\vec q_k^m\}_{k=1}^{J_{\Omega}}$ it takes the value
\begin{equation}
\label{eqn:ddensity}
\left. \rho^m\right|_{\vec x=\vec q_k^m}=\left\{
\begin{array}{ll}
\rho_1, &  \mbox{if}\ \vec q_k^m\in\bar{\Omega}_1^m\backslash\Gamma^m, \vspace{0.15cm}\\
\frac{1}{2}(\rho_1+\rho_2), & \mbox{if}\ \vec q_k^m\in\Gamma^m,\vspace{0.15cm} \\
\rho_2, & \mbox{if}\ \vec q_k^m\in\bar{\Omega}_2^m\backslash\Gamma^m.
\end{array}\right.
\end{equation}
We note that the density $\rho^m$ is a continuous function instead of
a piecewise constant function. This facilitates the interpolation of $\rho^m$ 
from the mesh $\mathcal{T}^m$ to the mesh $\mathcal{T}^{m+1}$ which is required in the 
numerical method.

Let $\mathbb{U}^m$ and $\hat{\mathbb{P}}^m$ denote the finite element spaces 
for the numerical solution for the velocity and pressure, respectively. 
We use the following two pairs of elements for 
$\left(\mathbb{U}^m,~\hat{\mathbb{P}}^m\right)$,
\begin{subequations}
\begin{align}
&{\rm P2-P0}:\ \left(\mathbb{U}^m,~\hat{\mathbb{P}}^m\right)=
\left([S_2^m]^2\cap\mathbb{U},~  S_0^m\cap \hat{\mathbb{P}} \right),\\
&{\rm P2-(P1+P0)}:\ \left(\mathbb{U}^m,~\hat{\mathbb{P}}^m\right)=
\left([S_2^m]^2\cap\mathbb{U},~  (S_1^m+S_0^m)\cap \hat{\mathbb{P}}   \right),
\end{align}
\end{subequations}
where $\mathbb{U}$ and $\hat{\mathbb{P}}$ are defined in \eqref{eqn:USpace} 
and \eqref{eq:P}, respectively.  
These two choices satisfy the inf-sup stability condition~\cite{Barrett15stable, Agnese16}, 
\begin{equation}
\inf_{\varphi\in  \hat{\mathbb{P}}^m} 
\sup_{\vec 0\neq\boldsymbol{\omega}\in \mathbb{U}^m}
\frac{\left(\varphi, \nabla\cdot\boldsymbol{\omega}\right)}
{\norm{\varphi}_0\norm{\boldsymbol{\omega}}_1}\geq C_0>0,
\label{eqn:LBB}
\end{equation}
where $\norm{\cdot}_0$ and $\norm{\cdot}_1$ denote the $L^2$ and $H^1$-norm 
on $\Omega$ respectively, and $C_0$ is a constant.
The finite element spaces for the pressure can catch the discontinuity 
of the pressure across the fluid interface.

We use $V^h\times V_0^h$ and $V^h$ as the finite element space for
the fluid interface and its curvature, respectively.
The finite element method is given as follows.
Let $\Gamma^0:=\vec X^0(\cdot )\in V^h\times V_0^h$ and $\mathcal{T}^0$ 
be the discretization of the initial interface $\Gamma(0)$ and the triangulation of
the domain $\Omega(0)$, respectively, and $\vec u^0=I_2^0\vec u_0\in \mathbb{U}^0$ be
the discretization of the initial fluid velocity $\vec u_0$. 
For $m \ge 0$, find $\vec u^{m+1}\in \mathbb{U}^m$, $p^{m+1}\in\hat{\mathbb{P}}^m$, 
$\vec X^{m+1}\in V^h\times V_0^h$, and $\kappa^{m+1}\in V^h$ by solving
the linear system
\begin{subequations}
\begin{align}
\label{eqn:full1}
&\frac{1}{2}\Bigl[\Bigl(\frac{\rho^m\vec u^{m+1}-(I_1^m\rho^{m-1})
I_2^m\vec u^m}{\tau_m},~\boldsymbol{\omega}^h\Bigr)
+\Bigl(I_1^m\rho^{m-1}\frac{\vec u^{m+1}-I_2^m\vec u^m}
{\tau_m},~\boldsymbol{\omega}^h\Bigr) +\Bigl(\rho^m(I_2^m\vec u^m\cdot\nabla)
\vec u^{m+1},~\boldsymbol{\omega}^h\Bigr)\nonumber\\
&\qquad\qquad -\Bigl(\rho^m(I_2^m\vec u^m\cdot\nabla)\boldsymbol{\omega}^h,~\vec u^{m+1}
\Bigr)\Bigr] - \Bigl(p^{m+1},~\nabla\cdot\boldsymbol{\omega}^h\Bigr)
+\frac{2}{Re} \Bigl(\eta^m D(\vec u^{m+1}),~D(\boldsymbol{\omega}^h)\Bigr)\nonumber\\
&\qquad\qquad -\frac{1}{We}\Bigl(\kappa^{m+1}\vec n^m,~\boldsymbol{\omega}^h
\Bigr)_{\Gamma^m}+ \frac{1}{Re\cdot l_s}\Bigr(\beta^m\,u_s^{m+1},
~\omega_s^h\Bigr)_{\Gamma_1^m\cup\Gamma_2^m}=0,\quad\forall\boldsymbol{\omega}^h
\in \mathbb{U}^m,\\[0.5em]
\label{eqn:full2}
&\qquad\qquad\qquad\qquad\qquad\Bigl(\nabla\cdot\vec u^{m+1},~q^h\Bigr)=0,
        \qquad\forall q^h\in \hat{\mathbb{P}}^m,\\[0.5em]
\label{eqn:full3}
&\qquad\qquad \Bigl(\frac{\vec X^{m+1}-\vec X^m}{\tau_m}\cdot\vec n^m,
~\psi^h\Bigr)_{\Gamma^m}^h - \Bigl(\vec u^{m+1}\cdot\vec n^m,~\psi^h\Bigr)
_{\Gamma^m}=0,\quad\forall \psi^h\in V^h,\\[0.5em]
&\Bigl(\kappa^{m+1}\,\vec n^m,~\boldsymbol{g}^h\Bigr)_{\Gamma^m}^h
+\Bigl(\partial_s\vec X^{m+1},~\partial_s\boldsymbol{g}^h\Bigr)_{\Gamma^m}
 -\cos\theta_Y \left[g_1^h(1) - g_1^h(0)\right]\nonumber\\
&\qquad\qquad\qquad 
+\frac{\beta^* Ca}{\tau_m}\Bigl[\left(x^{m+1}_r- x^m_r\right)g_1^h(1)
+\left(x^{m+1}_l-x_l^m\right) g_1^h(0)\Bigr]=0,
\quad\forall\boldsymbol{g}^h\in V^h\times V_0^h,
\label{eqn:full4}
\end{align}
\end{subequations}
where $\boldsymbol{g}^h=(g^h_1,~g^h_2)$, 
$\omega_s^h=\boldsymbol{\omega}^h\cdot\vec t_w$, $u_s^{m+1}=\vec u^{m+1} \cdot \vec t_w$, 
and $x_l^m:=X^m|_{\alpha=0}$ and $x_r^m=X^m|_{\alpha=1}$ denote the left and right 
contact points of $\Gamma^{m}$, respectively. For $f\in V^h$, $\partial_s f := \frac{1}{|\partial_\alpha\vec X^m|}\partial_\alpha f$.
At the first step, we set $\rho^{-1} = \rho^0$. 

In the above scheme, $\vec u^m$ and $\rho^{m-1}$ are both obtained on the mesh 
$\mathcal{T}^{m-1}$, and then used to compute the solutions 
$(\vec u^{m+1},\ p^{m+1},\ \vec X^{m+1},\ \kappa^{m+1})$ on the new mesh $\mathcal{T}^m$.
Therefore, we need to perform interpolations to obtain their values on the new mesh. 
The operators $I_1^m$ and $I_2^m$ are for this purpose. They denote the linear and 
quadratic interpolations from $\mathcal{T}^{m-1}$ to $\mathcal{T}^{m}$, respectively.

The numerical scheme is an extension of the earlier work by Barrett et. al. 
\cite{Barrett15stable} 
to systems with the moving contact lines. 
We note that the special treatment of the inertia term in Eq.~\eqref{eqn:immediate1}
is to maintain the discrete stability for the fluid kinetic energy. 
Another remark is on the disretization of the temporal derivative 
$\frac{d}{dt}\bigl(\rho\vec u,~\boldsymbol{\omega}\bigr)$,
\begin{equation}
\frac{d}{dt}\Bigl(\rho\vec u,~\boldsymbol{\omega}\Bigr)\approx \frac{1}{\tau_m}\Big[\Bigl(\rho^m\vec u^{m+1},~\boldsymbol{\omega^h}\Bigr)-\Bigl((I_1^m\rho^{m-1})(I_2^m\vec u^m),~\boldsymbol{\omega}^h\Bigr)\Bigr].
\end{equation}
The density $\rho^{m+1}$ depends on the mesh $\mathcal{T}^{m+1}$ 
(see the definition in Eq.~\eqref{eqn:ddensity}), thus is 
unknown before the interface $\Gamma^{m+1}$ is computed. Therefore, 
in the above discretization we avoided using $\rho^{m+1}$ 
by lagging the density by one time step. This yields a linear system for the 
solutions at $t=t_{m+1}$. 

The numerical scheme is a combination of the finite element method 
for the incompressible Navier-Stokes equations and the parametric finite element method 
for the interface evolution. The curvature is introduced as a new variable and treated implicitly in the scheme. This helps to yield the discrete stability for the interfacial energy as discussed next. The different numerical quadratures have been utilized to approximate the inner product over $\Gamma^m$, and the approximation by the mass-lumped norm is essential to the property of the equal mesh distribution, which has been discussed in detail in \cite{Barrett07}.

\subsection{Properties of the FEM}  

Next we show that the numerical method \eqref{eqn:full1} - \eqref{eqn:full4} 
yields a unique solution (Theorem \ref{th:unique1}), and is energy stable
(Theorem \ref{th:s}).

\begin{thm}[Well-posedness]
\label{th:unique1}
Let $(\mathbb{U}^m,~\hat{\mathbb{P}}^m)$ satisfy the inf-sup stability condition 
\eqref{eqn:LBB}, the interface $\vec X^m(\cdot)$ satisfy the conditions in 
\eqref{eqn:assumption}. Then the numerical method \eqref{eqn:full1}-\eqref{eqn:full4} 
admits a unique solution.
\begin{proof}
It suffices to show that the corresponding homogeneous system has only zero solution.
Thus we consider solving the following homogeneous system for 
$\big(\vec u^h,~p^h,~\vec X^h,~\kappa^h\big)\in
\big(\mathbb{U}^m,~\hat{\mathbb{P}}^m,~V^h\times V_0^h,~V^h\big)$,
\begin{subequations}
\begin{align}
\label{eqn:homoge1}
&\frac{1}{2}\Bigl[\Bigl(\frac{(\rho^m+I_1^m\rho^{m-1})\vec u^h}{\tau_m},
~\boldsymbol{\omega}^h\Bigr) +\Bigl(\rho^m(I_2^m\vec u^m\cdot\nabla)
\vec u^h,~\boldsymbol{\omega}^h\Bigr)-\Bigl(\rho^m(I_2^m\vec u^m\cdot\nabla)
\boldsymbol{\omega}^h,~\vec u^h\Bigr)\Bigr]\nonumber\\
&\qquad\qquad-\Bigl(p^h,~\nabla\cdot\boldsymbol{\omega}^h\Bigr)
+\frac{2}{Re}\,\Bigl(\eta^m D(\vec u^h),~D(\boldsymbol{\omega}^h)\Bigr) 
-\frac{1}{We}\Bigl(\kappa^h\,\vec n^m,~\boldsymbol{\omega}^h\Bigr)_{\Gamma^m} 
\nonumber\\
&\qquad\qquad\qquad\qquad\hspace{2.5cm} 
+\frac{1}{Re\cdot l_s}\Bigr(\beta^m\,u_s^h,~\omega_s^h\Bigr)_
 {\Gamma_1^m\cup\Gamma_2^m}=0,\quad\forall\boldsymbol{\omega}^h\in \mathbb{U}^m,\\[0.5em]
\label{eqn:homoge2}
&\qquad\qquad\qquad\qquad\qquad\qquad
\Bigl(\nabla\cdot\vec u^h,~q^h\Bigr)=0,\qquad\forall 
q^h\in \hat{\mathbb{P}}^m,\\[0.5em]
\label{eqn:homoge3}
&\qquad\qquad\qquad\Bigl(\frac{\vec X^h}{\tau_m}\cdot\vec n^m,~\psi^h\Bigr)
_{\Gamma^m}^h - \Bigl(\vec u^h\cdot\vec n^m,~\psi^h\Bigr)_{\Gamma^m}=0,
\quad\forall \psi^h\in V^h,\\[0.5em]
&\Bigl(\kappa^h\,\vec n^m,~\boldsymbol{g}^h\Bigr)_{\Gamma^m}^h
+\Bigl(\partial_s\vec X^h,~\partial_s\boldsymbol{g}^h\Bigr)_{\Gamma^m}
+\frac{\beta^*Ca}{\tau_m}\Bigl[x_r^h\,g_1^h(1)+x_l^h\,g_1^h(0)\Bigr]=0,
\quad\forall\boldsymbol{g}^h\in V^h\times V_0^h,
\label{eqn:homoge4}
\end{align}
\end{subequations}
where $\vec X^h=(X^h,~Y^h)$, $u_s^h = \vec u^h\cdot\vec t_w$, and $x_l^h:=X^h|_{\alpha=0}$ and $x_r^h:=X^h|_{\alpha=1}$.

Setting $\boldsymbol{\omega}^h=\vec u^h$, $q^h = p^h$, 
$\psi^h=\frac{1}{We}\kappa^h$ and $\boldsymbol{g}^h = \frac{1}{We}\vec X^h$, then
combinning these equations yields
\begin{align}
& \frac{1}{2}\Bigl((\rho^m+I_1^m\rho^{m-1})\vec u^h,~\vec u^h\Bigr) 
+ \frac{2\tau_m}{Re}\Bigl(\eta^m\,D(\vec u^h),~D(\vec u^h)\Bigr) 
+ \frac{\tau_m}{Re\cdot l_s}\Bigl(\beta^m\,u_s^h,~u_s^h\Bigr)
_{\Gamma_1^m\cup\Gamma_2^m} \nonumber\\
& \qquad\qquad\hspace{2cm}
+ \frac{1}{We}\Bigl(\partial_s\vec X^h,~\partial_s\vec X^h
\Bigr)_{\Gamma^m}+ \frac{\beta^{*}}{Re\cdot \tau_m}[(x_r^h)^2+(x_l^h)^2]=0.
\end{align}
By Korn's inequality, we have
\begin{equation}
\norm{\vec u^h}_1\leq C\Bigl[\frac{1}{2}\Bigl((\rho^m+I_1^m\rho^{m-1})\vec u^h,~\vec u^h\Bigr) 
+ \frac{2\tau_m}{Re}\Bigl(\eta^m\,D(\vec u^h),~D(\vec u^h)\Bigr)\Bigr ]\leq 0,
\end{equation}
we immediately obtain $\vec u^h=\vec 0$.
By noting $x_r^h=x_l^h=0$,  we also have $\vec X^h=\vec 0$.
Next, by substituting $\vec X^h=\vec 0$ into Eq.~\eqref{eqn:homoge4}, we obtain 
\begin{equation}
\Bigl(\kappa^h\,\vec n^m,~\boldsymbol{g}^h\Bigr)_{\Gamma^m}^h=0,\qquad 
\forall\boldsymbol{g}^h\in V^h\times V_0^h .
\end{equation}
Choosing the test function $\boldsymbol{g}^h$ such that 
\begin{equation}
\left. \boldsymbol{g}^h\right|_{\alpha_j} =\left\{
\begin{array}{l}
-\left[\vec X^{m}(\alpha_{j+1}) - \vec X^m(\alpha_{j-1})\right]^\perp \kappa^h(\alpha_j),
\quad 1\leq j\leq J_{_\Gamma}-1, \vspace{0.15cm}\\
\left(n_1^{m,1}\kappa^h(\alpha_j), ~0\right), \quad j = 0,\vspace{0.15cm} \\
\left(n_{J_{_\Gamma}}^{m,1}\kappa^h(\alpha_j),~0\right),\quad j = J_{_\Gamma},
\end{array} \right.
\end{equation}
By the assumptions in \eqref{eqn:assumption} and the norm in \eqref{eqn:massnorm}, 
we obtain $\kappa^h(\alpha_j)=0,\;\forall 0\leq j\leq J_{_\Gamma}$, 
which implies $\kappa^h = 0$.  We then substitute $\vec u^h = \vec 0$ and $\kappa^h=0$ 
into Eq.~\eqref{eqn:homoge1} and obtain
 \begin{equation}
 \left(p^h,~\nabla\cdot\boldsymbol{\omega}^h\right)=0,
\qquad\forall \boldsymbol{\omega}^h\in \mathbb{U}^m.
 \end{equation}
Using the stability condition in Eq.~\eqref{eqn:LBB}, we consequently 
obtain $p^h=0$. This shows that the homogeneous linear system 
\eqref{eqn:homoge1} - \eqref{eqn:homoge4} has only the zero solution.  
Thus, the numerical scheme \eqref{eqn:full1}-\eqref{eqn:full4}
admits a unique solution.
\end{proof}
\end{thm}

We next show that the numerical scheme satisfies a stability bound 
in terms of a discrete energy corresponding to 
Eq.~\eqref{eqn:dimenenergy}.

\begin{thm}[Stability bound]   \label{th:s}
Let $\left(\vec u^{m+1},~p^{m+1},~\vec X^{m+1},~\kappa^{m+1}\right)$ 
be the solution to the numerical scheme 
\eqref{eqn:full1}-\eqref{eqn:full4}. Then the following stability bound holds
\begin{eqnarray}
&&\mathcal{E}(\rho^m,\vec u^{m+1},\Gamma^{m+1})+\frac{1}{2}
\norm{\sqrt{I_1^m\rho^{m-1}}(\vec u^{m+1}-I_2^m\vec u^m)}_0^2
+\frac{2\tau_m}{Re}\norm{\sqrt{\eta^m}D(\vec u^{m+1})}_0^2\nonumber\\
&& \qquad+ \frac{\tau_m}{Re\cdot l_s}\Bigl(\beta^m\,u_s^{m+1},~u_s^{m+1}\Bigr)
_{\Gamma^m_1\cup\Gamma_2^m} +\frac{\beta^*}{Re\cdot \tau_m}
\Bigl[\bigl(x^{m+1}_r-x^m_r\bigr)^2+\bigl(x^{m+1}_l-x^m_l\bigr)^2\Bigr]\nonumber\\[0.5em]
&&\qquad\qquad\qquad\qquad\qquad\qquad\leq
\mathcal{E}(I_1^m\rho^{m-1},I_2^m\vec u^m,\Gamma^m),
\label{eqn:energybounds}
\end{eqnarray}
where $\mathcal{E}(\rho,\vec u,\Gamma):=\frac{1}{2}(\rho\vec u,~\vec u) 
-\frac{\cos\theta_Y}{We}|\Gamma_1| + \frac{1}{We}|\Gamma|$ is the total 
energy of the system.
%
\begin{proof}
Setting $\boldsymbol{\omega}^h=\vec u^{m+1}$, $q^h = p^{m+1}$, 
$\psi^h = \frac{1}{We}\,\kappa^{m+1}$ and 
$\boldsymbol{g}^h = \frac{1}{We\cdot\tau_m}(\vec X^{m+1}-\vec X^m)$ in Eqs.
\eqref{eqn:full1}-\eqref{eqn:full4}, 
then combining these equations  yields
\begin{align}
&\frac{1}{2\tau_m }\left[\Bigl(\rho^m\vec u^{m+1}-I_1^m\rho^{m-1}I_2^m\vec u^m,
~\vec u^{m+1}\Bigr)+\Bigl(I_1^m\rho^{m-1}\left(\vec u^{m+1}-I_2^m\vec u^m\right),
~\vec u^{m+1}\Bigr)\right]   \nonumber\\
& \qquad +\,\frac{2}{Re}\,\Bigl(\eta^m D(\vec u^{m+1}),~D(\vec u^{m+1})\Bigr)
+\frac{1}{Re\cdot l_s}\left(\beta^m\,u_s^{m+1},~u_s^{m+1}\right)
_{\Gamma_1^m\cup\Gamma_2^m} \nonumber\\
& \qquad +\frac{1}{We\cdot\tau_m}\Bigl(\partial_s\vec X^{m+1},
~\partial_s(\vec X^{m+1}-\vec X^m)\Bigr)_{\Gamma^m}
-\frac{\cos\theta_Y}{We\cdot\tau_m}\Bigl[(x_r^{m+1}-x_l^{m+1})-(x_r^m-x_l^m)\Bigr]
\nonumber\\
&  \qquad
+\frac{\beta^*}{Re\cdot(\tau_m)^2} \Bigl[\bigl(x_r^{m+1}-x_r^m\bigr)^2
+\bigl(x_l^{m+1}-x_l^m\bigr)^2\Bigr]=0.
\label{eqn:energybound1}
\end{align}
It is easy to see that the following equality/inequality holds:
\begin{align}
&\Bigl(\rho^m\vec u^{m+1}-I_1^m\rho^{m-1}I_2^m\vec u^m,~\vec u^{m+1}\Bigr)
 +\Bigl(I_1^m\rho^{m-1}\left(\vec u^{m+1}-I_2^m\vec u^m\right),~\vec u^{m+1}\Bigr)
\nonumber\\
 =&\Bigl(\rho^m\,\vec u^{m+1},~\vec u^{m+1}\Bigr)
-\Bigl(I_1^m\rho^{m-1}\,I_2^m\vec u^{m},~I_2^m\vec u^{m}\Bigr)
+ \Bigl(I_1^m\rho^{m-1}(\vec u^{m+1}-I_2^m\vec u^m),
~\vec u^{m+1}-I_2^m\vec u^m\Bigr),   \label{eqn:energybound2}
\end{align}
\begin{align}
\Bigl(\partial_s\vec X^{m+1},~\partial_s(\vec X^{m+1}-\vec X^m)\Bigr)_{\Gamma^m}&\geq\;\frac{1}{2}\Bigl(|\partial_s\vec X^{m+1}|^2-|\partial_s\vec X^m|^2,~1\Bigr)_{\Gamma^m}\nonumber\\
&\geq\;\Bigl(|\partial_s\vec X^{m+1}|-1,~1\Bigr)_{\Gamma^m}
= |\Gamma^{m+1}|-|\Gamma^m|,
\label{eqn:energybound4}
\end{align}
where we have used $a(a-b)\geq \frac{1}{2}(a^2-b^2)$ and 
$\frac{a^2-1}{2}\geq |a|-1$ in Eq. \eqref{eqn:energybound4}.
Using Eqs. \eqref{eqn:energybound2} - \eqref{eqn:energybound4} 
in Eq. \eqref{eqn:energybound1} and noting $x_r^{m+1}-x_l^{m+1} = |\Gamma_1^{m+1}|$ and 
$x_r^m-x_l^m =|\Gamma_1^m|$, we immediately obtain Eq.~\eqref{eqn:energybounds}.
\end{proof}
\end{thm}
Eq.~\eqref{eqn:energybounds} gives a bound for the energy 
 $\mathcal{E}(\rho^m,\vec u^{m+1},\Gamma^{m+1})$ of the discrete system
at $t=t_{m+1}$ in terms of the energy 
$\mathcal{E}(I_1^m\rho^{m-1},I_2^m\vec u^m,\Gamma^m)$, where 
$I_1^m\rho^{m-1}$ and $I_2^m\vec u^m$ are interpolations of $\rho^{m-1}$ and 
$\vec u^m$ from $\mathcal{T}^{m-1}$ to $\mathcal{T}^m$, respectively.
Note that this does not imply energy dissipation in the whole time domain, i.e.
$\mathcal{E}(\rho^m,\vec u^{m+1},\Gamma^{m+1})\leq \mathcal{E}
(\rho^{m-1},\vec u^m,\Gamma^m)$, due to the interpolation errors.
Nevertheless, we did observe the decay of the energy in numerical simulations,
which will be shown in section \ref{sec:results}.

\subsection{The moving mesh}  \label{subsec:mesh}

The fitted mesh is generated using a moving meshg method. At the $m$th time step 
$(m\ge 0)$, a new mesh $\mathcal{T}^{m+1}=\bigcup_{j=1}^N\bar{o}_j^{m+1}$ is obtained 
by adapting the mesh at the previous time step so that it fits to
the newly obtained interface  $\Gamma^{m+1}$, i.e.
\begin{equation}
\Gamma^{m+1}\subset\bigcup_{j=1}^N\partial o_j^{m+1}. 
\end{equation}
Specifically, suppose we have solved for $\vec X^{m+1}$ on $\mathcal{T}^m$.
This gives $\Gamma^{m+1}$, the numerical solution for the interface at $t=t_{m+1}$. 
Then we construct the new mesh $\mathcal{T}^{m+1}$ based on
 $\mathcal{T}^m$, where the mesh connectivity and topology remain unchanged.
This is achieved by updating the vertices of the triangular mesh as
\begin{equation}  \label{eq:qk}
\vec q_k^{m+1} = \vec q_k^m + \boldsymbol{\eta}|_{\vec q_k^m},\quad k=1,\cdots, N
\end{equation}
where $\boldsymbol{\eta}=(\eta^1,~\eta^2)\in [S_1^m]^2$ is the displacement vector.
The displacement of the vertices on the boundary $\Gamma_1^m\cup\Gamma_2^m$ 
is $\boldsymbol{\eta}=(\eta^1(x),~0)$, where $\eta^1(x)$ is 
the piecewise linear function taking the values 0, $\Delta x_l^m:=x_l^{m+1}-x_l^m$, 
$\Delta x_r^m:=x_r^{m+1}-x_r^m$ and 0 at 
$x=-L_x,\ x_l^m,\ x_r^m$ and $L_x$, respectively, i.e.
\begin{equation}   \label{eq:eta1}
\eta^1(x)= \left\{
\begin{array}{l}
\frac{\Delta x_l^m (x+L_x)}{x_l^m+L_x},\quad -L_x\le x<x_l^m, \vspace{0.15cm}\\
\frac{\Delta x_l^m(x-x_r^m)}{x_l^m-x_r^m} +\frac{\Delta x_r^m(x-x_l^m)}{x_r^m-x_l^m},
\quad x_l^m\le x\le x_r^m,\vspace{0.15cm} \\
\frac{\Delta x_r^m(x-L_x)}{x_r^m-L_x},\quad x_r^m<x\le L_x.
\end{array}\right.
\end{equation}
The displacements of the internal vertices are obtained by solving the equation
\cite{Masud97space, Liu16}
\begin{equation} \label{eqn:elastic}
\nabla\cdot \left[\lambda(\vec x)\left(\nabla\boldsymbol{\eta} 
+ (\nabla\boldsymbol{\eta})^T+(\nabla\cdot\boldsymbol{\eta})\mathbf{I}\right)\right]=\vec 0
\end{equation}
on $\mathcal{T}^m$ with $\mathcal{P}^1$ Lagrange element, 
with the boundary conditions $\boldsymbol{\eta} = \vec X^{m+1}- \vec X^m$
on $\Gamma^m$, $\boldsymbol{\eta} = \vec 0$ on $\Gamma_3^m\cup\Gamma_4^m$ and 
$\boldsymbol{\eta} = (\eta^1,0)$ on $\Gamma_1^m\cup\Gamma_2^m$, where 
$\eta^1$ is given in Eq. \eqref{eq:eta1}.
Here $\lambda(\vec x)$ is defined as
\begin{equation}
\lambda(\vec x)|_{o_i^m}:=1 + \frac{\max_{j=1}^N|o_j^m|-\min_{j=1}^N|o_j^m|}{|o_i^m|},
\end{equation}
and it is used to limit the distortion of small elements.

Instead of the moving mesh approach, one may use fixed mesh in the discretization
\cite{Barrett15stable}.
This avoids the interpolation between the meshes, thus
the global energy stability can be achieved. 
The drawback is that, at each time step,
one needs to determine the intersections of  the line segments of the interface 
with the triangles, since the computational mesh for the moving interface 
is decoupled from the mesh for the Naiver-Stokes equation. This is rather complicated, 
especially in high dimensions. 
Moreover, additional work needs to be done to capture the pressure jump
across the interface and to ensure the area conservation in the unfitted mesh approach. 

\vspace{0.5cm}
The overall procedure of the numerical method is summarised as follows. 
Given the initial velocity $\vec u^0=I_2^m\vec u_0$ and the 
interface $\Gamma^0$, let  $\mathcal{T}^0$ be a triangulation of $\Omega$,
 $\rho^{-1}=\rho^0$, and $m=0$. Then
\begin{itemize}
\item[(1)] Solve the linear system Eq.~\eqref{eqn:full1}-\eqref{eqn:full4} 
on $\mathcal{T}^m$ for $\vec u^{m+1}$, $p^{m+1}$, $\Gamma^{m+1}:=\vec X^{m+1}$
and $\kappa^{m+1}$;
\item[(2)] Solve Eq. \eqref{eqn:elastic} on $\mathcal{T}^m$ for the displacement vector 
$\boldsymbol{\eta}$, and construct the new mesh $\mathcal{T}^{m+1}$ according to
Eq. \eqref{eq:qk};
\item[(3)] Perform interpolations from $\mathcal{T}^m$ to $\mathcal{T}^{m+1}$
to obtain $I_2^{m+1}\vec u^{m+1}$ and $I_1^{m+1}\rho^{m}$, and go to step (1) with m= m+1. 
\end{itemize}

\section{Numerical results} \label{sec:results}

In this section, we present the convergence test and some numerical examples
for the proposed FEM method. In the simulations, 
we use $Re=10$, $l_s=0.1$ unless otherwise stated. Other parameters will be specified later.
The initial velocity of the fluids is $\vec u_0=\vec 0$.

\subsection{Convergence test}  \label{subsec:con}

\begin{figure}[t]
\centering
\includegraphics[width=0.75\textwidth]{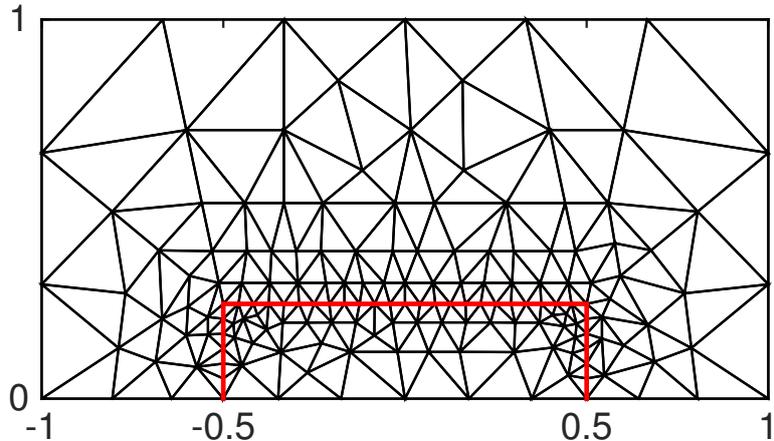}
\caption{The initial computational mesh used in the convergence test 
with $J_{_\Gamma}=36$, $J_{\Omega}=166$ and $N=306$.
The red line represents the fluid interface.} \label{fig:initialmesh}
\end{figure}

We first investigate the convergence of the proposed numerical method 
by carrying out simulations with different mesh sizes and time steps. 
The computational domain is $\Omega=[-1,1]\times[0,1]$ (i.e. $L_x=L_y=1$). 
Initially, the region occupied by fluid $1$ is  the rectangle 
$\Omega_1(0)=[-0.5, 0.5]\times[0,0.25]$. The parameters are chosen as 
$\rho_1=0.1$, $\beta_1=0.1$, $\eta_1=10$, $\beta^*=0.1$, $\theta_Y=2\pi/3$
and $Ca = 0.01$.

Let $\vec X^m(\alpha)$ be the numerical solution for the interface at $t=t_m\ (m\ge 0)$ 
obtained with the time step $\tau$ and mesh size $h=1/J_\Gamma$. 
We define the approximate solution in any time interval $t_{m}\leq t<t_{m+1}$ 
using the linear interpolation:
%
\begin{equation}
 \vec{X}_{h,\tau}(\alpha,t)  =\frac{t-t_{m}}{\tau}\vec{X}^{m+1}(\alpha)
+\frac{t_{m+1}-t}{\tau}\vec{X}^{m}(\alpha).
\end{equation}
Then we measure the error of the numerical solution by comparing it with 
$\vec X_{\frac{h}{2},\frac{\tau}{4}}$, the numerical solution computed using refined
mesh and time step,
\begin{equation}\label{errorn}
e_{h,\tau}(t): 
=\max_{0\leq j\leq J_{_\Gamma}}\min_{\alpha\in [0,1]}
\left|\vec{X}_{h,\tau}(\alpha_j,t)-\vec{X}_{\frac{h}{2},\frac{\tau}{4}}(\alpha,t)\right|.
\end{equation} 
In Table.~\ref{tb:order1}, we report the error of the numerical solution 
at the three different times $t=0.2,\ 1.0,\ 4.0$ for the two choices of elements
P2-P0 and P2-(P1+P0), respectively. We observe that the error decreases with refined
mesh size and time step. However, the order of convergence is unstable. 
This is due to the accumulation of the errors induced 
in the interpolations of the density and velocity fields, 
which are carried out at each time step.

\begin{table}[t]
\centering
\def\temptablewidth{0.95\textwidth}
\vspace{0pt}
\caption{Error of the numerical solution and the rate of convergence 
for the fluid interface modelled using the Navier-Stokes equations. 
$h=1/J_\Gamma$ and $\tau$ are the mesh size in the discretization of the interface and
the time step, respectively, where $h_0=1/36$ and $\tau_0=0.01$. 
The numerical results are obtained using the P2-P0 elements (upper panel) and
the P2-(P1+P0) elements (lower panel).
}
{\rule{\temptablewidth}{1pt}}
\begin{tabular*}{\temptablewidth}{@{\extracolsep{\fill}}ccccccc}
$(h,\ \tau)$ 
&$e_{h,\tau}(t=0.2) $ & order &$e_{h,\tau}(t=1.0)$ & order &$e_{h,\tau}(t=4.0) $ & order  \\ \hline
$(h_0, \tau_0)$ 
& 5.86E-3 & - &5.03E-3 &-& 5.75E-3 &- \\ \hline
$(\frac{h_0}{2}, \frac{\tau_0}{2^2})$ 
& 1.97E-3 & 1.57 &1.07E-3 &2.23& 1.13E-3 &2.35 \\ \hline
$(\frac{h_0}{2^2}, \frac{\tau_0}{2^4})$ 
& 4.54E-4 & 2.12 &5.74E-4 &0.90& 7.09E-4 &0.67 
 \end{tabular*}
{\rule{\temptablewidth}{1pt}}
{\rule{\temptablewidth}{1pt}}
\begin{tabular*}{\temptablewidth}{@{\extracolsep{\fill}}ccccccc}
$(h,\ \tau)$ &$e_{h,\tau}(t=0.2) $ & order &$e_{h,\tau}(t=1.0)$ & order &$e_{h,\tau}(t=4.0) $ & order  \\ \hline
$(h_0, \tau_0)$ 
& 4.71E-3 & - &4.28E-3 &- &4.65E-3& - \\ \hline
$(\frac{h_0}{2}, \frac{\tau_0}{2^2})$ 
& 1.58E-3 & 1.58 &1.47E-3 &1.54& 1.74E-3 &1.42 \\ \hline
$(\frac{h_0}{2^2}, \frac{\tau_0}{2^4})$ 
& 4.25E-4 & 1.89 &6.49E-4 &1.18 & 8.18E-4 &1.09
 \end{tabular*}
{\rule{\temptablewidth}{1pt}}
\label{tb:order1}
\end{table}

\begin{figure}[tp]
\centering
\includegraphics[width=1.0\textwidth]{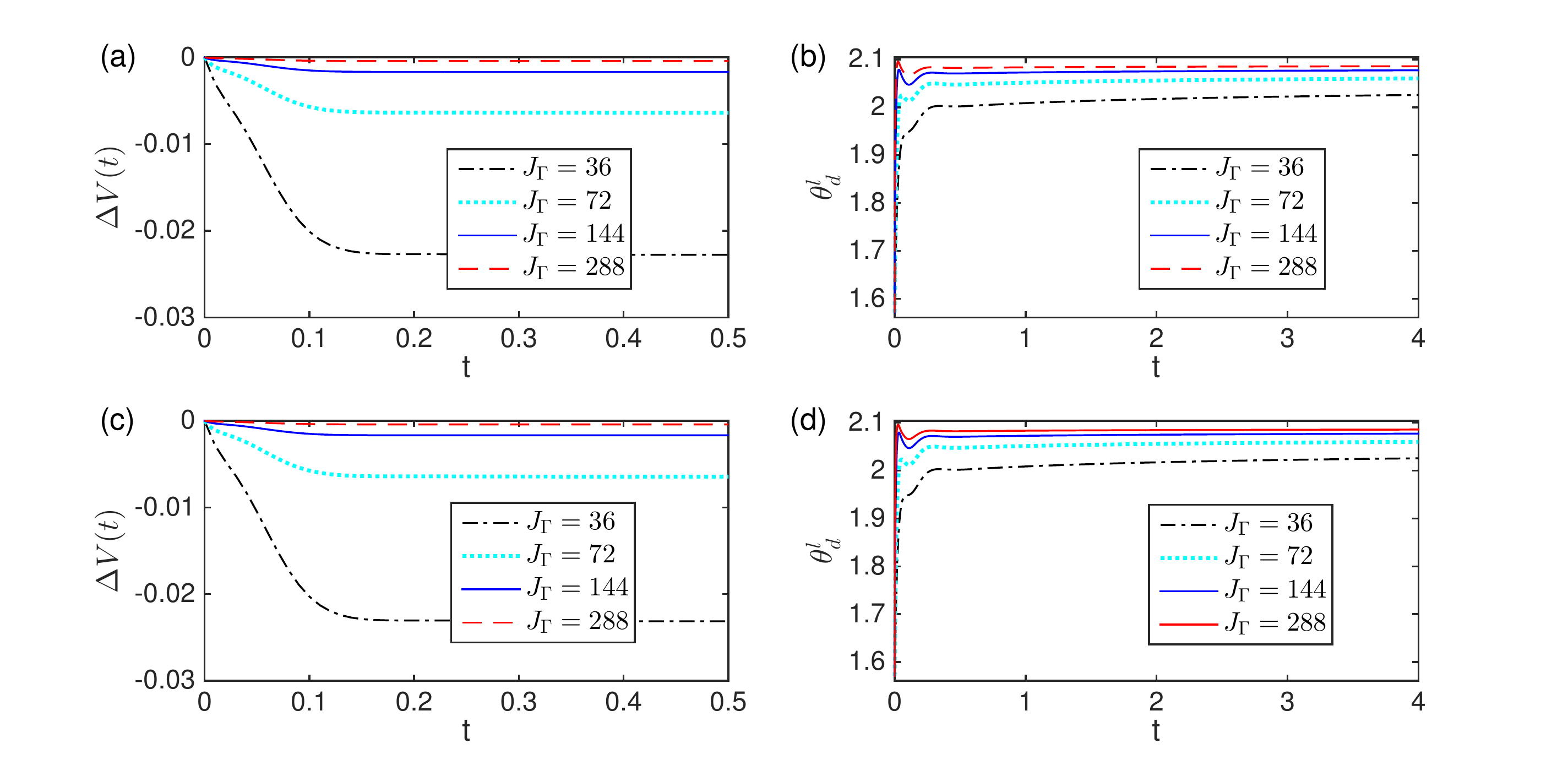}
\caption{The relative area change $\Delta V(t)$ of the droplet 
and the dynamic contact angle $\theta_d^l$ versus time for four different mesh sizes.
The numerical results are obtained using the P2-P0 elements (upper panels) 
and the P2-(P1+P0) elements (lower panels). }
\label{Fig:NVVolumeAngle}
\end{figure}

\begin{table}[tp]
\centering
\def\temptablewidth{0.95\textwidth}
\caption{Convergence rates of $\Delta V$, the relative area change of the droplet,
and $\theta_d^l$, the dynamic contact angle at $t=4$. $h$ and $\tau$ are the 
mesh size in the discretization of the interface and the time step, respectively,
where $h_0=1/36$ and $\tau_0=0.01$. The numerical results are obtained 
 using the P2-P0 elements (upper panel) and the P2-(P1+P0) elements (lower panel). }
{\rule{\temptablewidth}{1pt}}
\begin{tabular*}{\temptablewidth}{@{\extracolsep{\fill}}cccccc}
$(h,\ \tau)$  & $|\Delta V(t)|(t=4)$ & order &$|\theta_d^l(t)-\theta_Y|(t=4)$ & order  
\\ \hline
$(h_0, \tau_0)$ & 2.28E-2 & - &6.86E-2 &- \\ \hline
$(\frac{h_0}{2}, \frac{\tau_0}{2^2})$ & 6.38E-3 & 1.84 &3.41E-2 &1.01 \\ \hline
$(\frac{h_0}{2^2}, \frac{\tau_0}{2^4})$ &  1.68E-3
  & 1.93 &1.70E-2 &1.00 \\ \hline
$(\frac{h_0}{2^3}, \frac{\tau_0}{2^6})$ & 4.28E-4 & 1.97 &8.58E-3 &0.99 
 \end{tabular*}
{\rule{\temptablewidth}{1pt}}
{\rule{\temptablewidth}{1pt}}
\begin{tabular*}{\temptablewidth}{@{\extracolsep{\fill}}cccccc}
$(h,\ \tau)$ & $|\Delta V(t)|(t=4) $ & order &$|\theta_d^l(t)-\theta_Y|(t=4)$ & order   
\\ \hline 
$(h_0, \tau_0)$ & 2.31E-2 & - &6.86E-2 &- \\ \hline
$(\frac{h_0}{2}, \frac{\tau_0}{2^2})$ & 6.44E-3 & 1.84 &3.41E-2 &1.01\\ \hline
$(\frac{h_0}{2^2}, \frac{\tau_0}{2^4})$ & 1.68E-3 
& 1.94 &1.70E-2 &1.00\\\hline
$(\frac{h_0}{2^3}, \frac{\tau_0}{2^6})$ & 4.28E-4 & 1.97 &8.51E-3 &1.00
 \end{tabular*}
{\rule{\temptablewidth}{1pt}}
\label{tb:order2}
\end{table}

In Fig.~\ref{Fig:NVVolumeAngle}, we present the relative area change of the droplet
(left panels) and the dynamic contact angle (right panels) obtained 
using four different mesh sizes. The relative area change is defined as
\begin{equation}
\Delta V(t): = \frac{|\Omega_1(t)|-|\Omega_1(0)|}{|\Omega_1(0)|}.
\end{equation}
As can be seen from the numerical results, 
by refining the mesh the area loss is significantly 
reduced for both pairs of elements. We also observe the convergence of the dynamic 
contact angle as the mesh is refined. 

A more quantitative assessment for the area change and
the contact angle is provided in Table ~\ref{tb:order2}, 
where we show the area change and the convergence of contact angle 
to its equilibrium value $\theta_Y = 2\pi/3$ after the steady state is reached ($t=4$).
We observe that both errors decrease as the mesh is refined. 
The convergence order for $\Delta V$ is about $2$, and the convergence order 
for $|\theta_d^l- \theta_Y|$ is about $1$. 
The later can be understood as follows.
By choosing the test function $\boldsymbol{g}^h =(\phi_0(\alpha), ~0)$ in 
\eqref{eqn:full4}, where $\phi_0(\alpha)\in V^h$ is the piecewise
linear function taking the value 1 at $\alpha_0=0$ and 0 at other nodes (i.e.
the hat function at $\alpha_0$),
we obtain
\begin{equation}
\frac{1}{2}\kappa^{m+1}(0)n_1^{m,1}\left|\vec X^m(\alpha_1)-\vec X^m(\alpha_0)\right| 
- \Bigl(\frac{\partial_\alpha\vec X^{m+1}}{|\partial_\alpha\vec X^m|}\Bigr)\Big|_{\alpha=0} + \cos\theta_Y + \frac{\beta^* Ca}{\tau_m}
\left(x_l^{m+1} - x_l^m\right) = 0.
\end{equation}
At the steady state, we have 
$\vec X^{m+1}=\vec X^m$ and $\Bigl(\frac{\partial_\alpha\vec X^{m+1}}{|\partial_\alpha\vec X^m|}\Bigr)\Big|_{\alpha=0}= \partial_s\vec X^m\big|_{\alpha=0}=\cos\theta_d^{l,m}=\cos\theta_d^l$, thus, 
\begin{equation}
\cos\theta_d^l - \cos\theta_Y =  
\frac{1}{2}\kappa_h(0)n_{1,h}^{1}\left|\vec X_h(\alpha_1)-\vec X_h(\alpha_0)\right| 
= O(\kappa_h (0) h),
\label{eqn:angleerror}
\end{equation}
where the subscript $h$ denotes the numerical solution at the steady state. This explains the order of convergence for the contact angle 
shown in Table \ref{tb:order2}.

We note that in this example (and examples below), 
the parameter $\alpha$ is chosen as the normalized 
arc length of the initial interface $\Gamma(0)$. Thus the mesh points are evenly distributed
along $\Gamma(0)$. Since an implicit tangential velocity has been introduced for the interface evolution, and the mesh points tend to be uniformly distributed \cite{Barrett07, Bao17}, thus the quality of the mesh is well-preserved and no re-meshing
is needed in the computation.


\subsection{Numerical examples}

\begin{figure}[tph]
\centering
\includegraphics[width=0.95\textwidth]{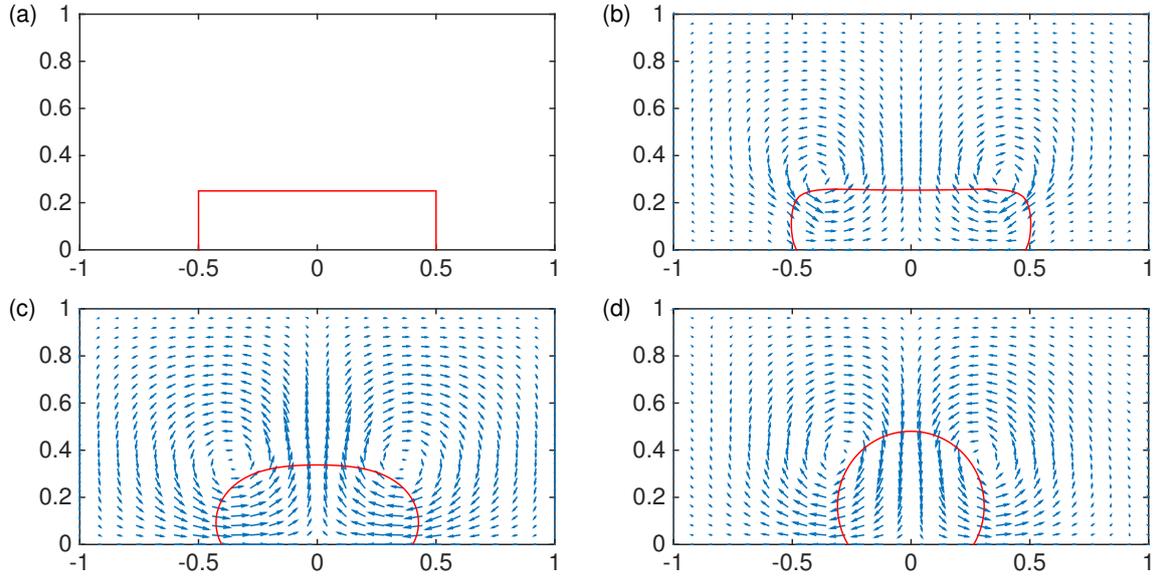}
\caption{Snapshots of the interface and the velocity field modeled 
by two-phase Navier-Stokes equations, where $\theta_Y=2\pi/3$. 
(a) $t=0$: $\max\norm{\vec u}_0=0$; (b) $t=0.1$: $\max\norm{\vec u}_0=0.231$; (c) $t=0.5$: $\max\norm{\vec u}_0=0.281$; (d) $t=2.0$: $\max\norm{\vec u}_0=0.012$.}
\label{Fig:Dewet}
\end{figure}

\begin{figure}[tph]
\centering
\includegraphics[width=0.95\textwidth]{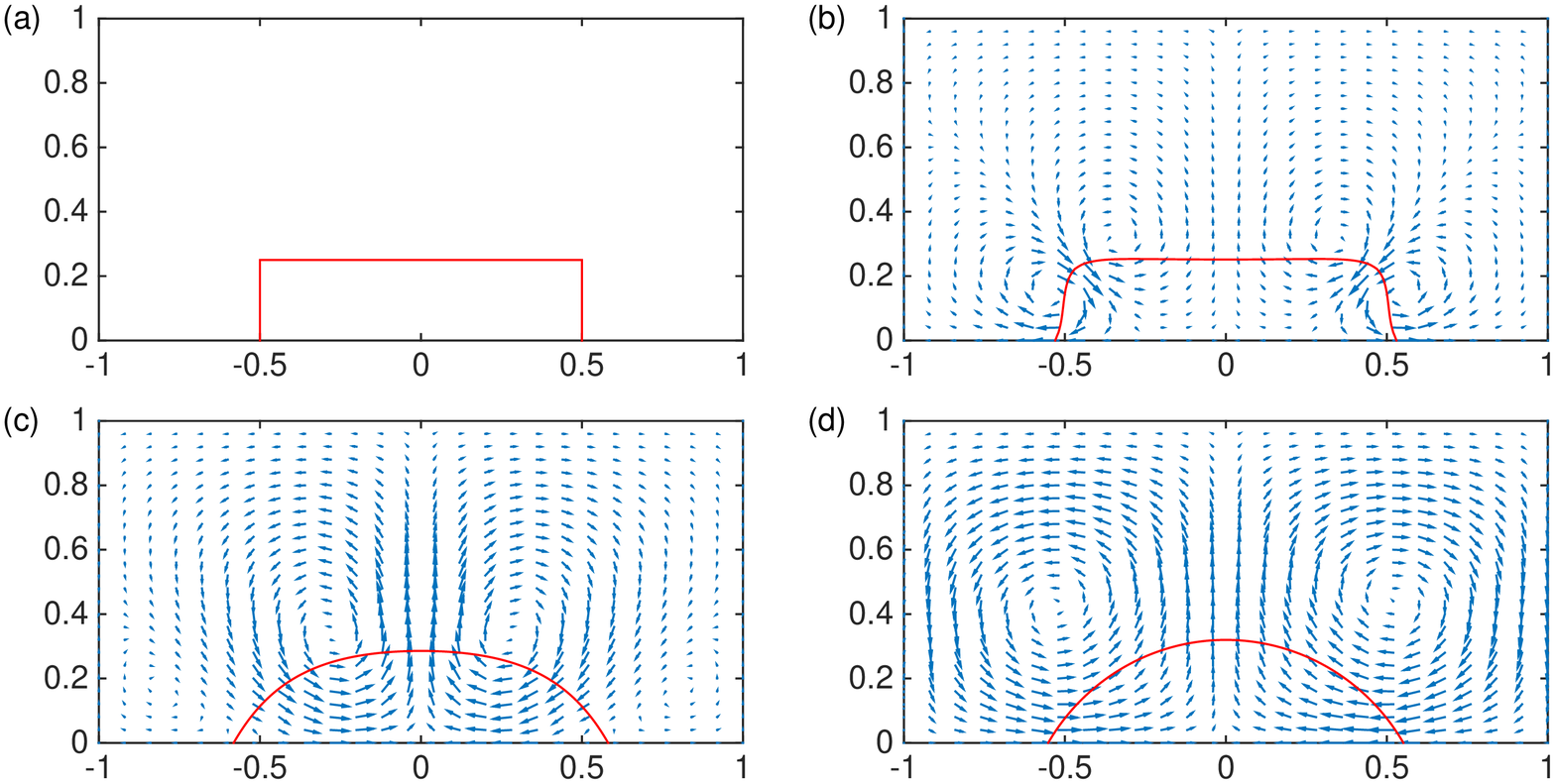}
\caption{Snapshots of the interface and the velocity field modeled 
by two-phase Navier-Stokes equations, where $\theta_Y=\pi/3$. 
(a) $t=0$: $\max\norm{\vec u}_0=0$; (b) $t=0.1$: $\max\norm{\vec u}_0=0.305$; (c) $t=0.5$: $\max\norm{\vec u}_0=0.075$; (d) $t=2.0$: $\max\norm{\vec u}_0=0.002$. }
\label{Fig:Wet}
\end{figure}

\begin{figure}[tph]
\centering
\includegraphics[width=0.95\textwidth]{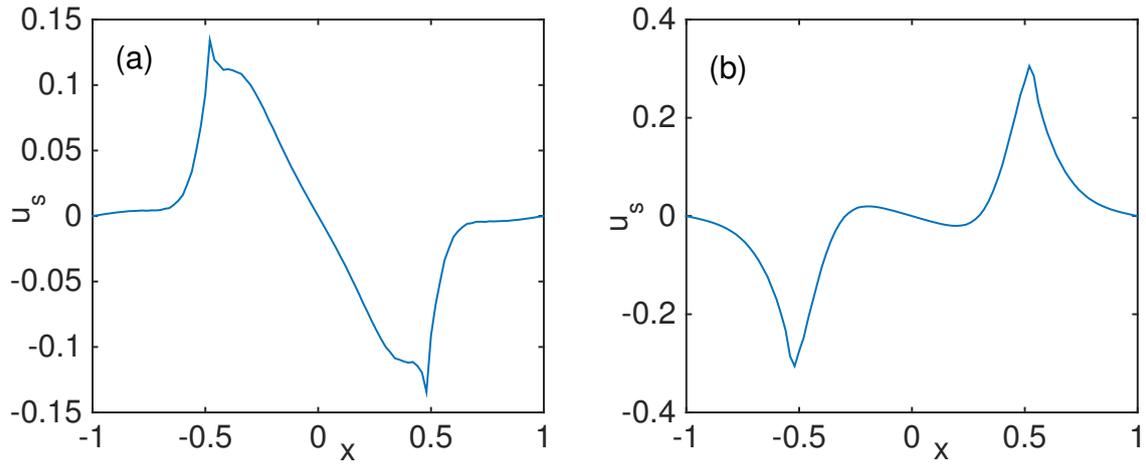}
\caption{The slip velocity $u_s=\vec u\cdot\vec t_w$ at $t=0.1$.
(a): the dewetting case with $\theta_Y=2\pi/3$; 
(b): the wetting case with $\theta_Y=\pi/3$.  }
\label{Fig:slipvelocity}
\end{figure}

\begin{figure}[tph]
\centering
\includegraphics[width=0.95\textwidth]{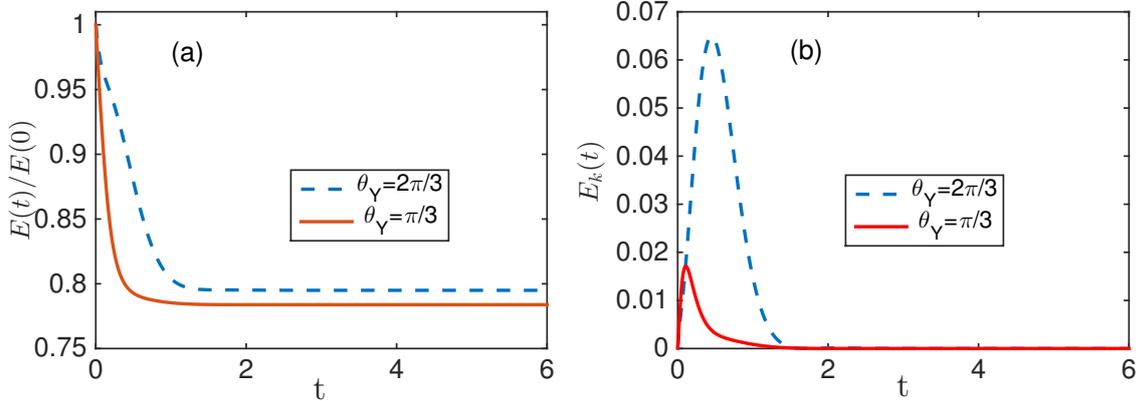}
\caption{The normalized energy $E(t)/E(0)$ (left panel) and the  kinetic energy 
$E_k(t):=\int_{\Omega}\frac{1}{2}\rho |\vec u|^2 d\mathcal{L}^2$ (right panel)
versus time. 
}
\label{Fig:EnergyAngle}
\end{figure}

Next we present two numerical examples. The first is similar to the one 
used in the convergence
test but with different parameters, and the second is the transport of a droplet on
solid substrate due to a surface tension gradient. 
The numerical results obtained using the P2-P0 elements and P2-(P1+P0) elements are 
indistinguishable in visualization, thus we will only present the results obtained using
 the P2-(P1+P0) elements.

\vspace{0.3cm}
\noindent {\it Example 1.} 
We first consider the evolution of a droplet on solid substrates
with different equilibrium contact angles: $\theta_Y=\frac{2\pi}{3}$ 
and $\theta_Y=\frac{\pi}{3}$.
The initial configuration of the droplet is given by a rectangle. The computational domain
is $\Omega=[-1, 1]\times[0, 1]$, which is discretized by 
the triangular mesh with $N=3348$ triangles and $J_\Omega=1716$ vertices;
the interface contains $J_\Gamma=120$ line segments. 
The time step is $\tau = 5\times 10^{-4}$.
Other parameters are chosen as 
$\rho_1=10$, $\beta_1 = 0.1$, $\eta_1 = 10$, $\beta^*=0.1$, and $Ca=0.1$.

Snapshots of the interface and the velocity field at several times 
are shown in Fig.~\ref{Fig:Dewet} and Fig.~\ref{Fig:Wet} for the two cases , respectively. 
In both cases, we can clearly observe the development of a pair of vortices 
in the velocity field associated with the evolution of the interface. 
In the dewetting case ($\theta_Y=2\pi/3$), inward velocities
are generated at the contact points due to the unbalanced Young stress, 
causing the contact points to retreat so that the contact angle converges to 
its equilibrium value. 
On the other hand, outward velocities are generated at the contact points 
in the wetting case ($\theta_Y=\pi/3$), which drives the droplet to spread on the substrate. 
The slip velocities along the substrate at time $t=0.1$ 
are shown in Fig.~\ref{Fig:slipvelocity}.
We can observe that the slip velocity takes the maximal value (in magnitude)
at the contact points in both cases. 

In Fig.~\ref{Fig:EnergyAngle}, we show the total and kinetic energies against time.
In particular, we observe the decay of the total energy in time.

\vspace{0.3cm}

\noindent {\it Example 2.} 
We next consider the migration of a droplet on a solid substrate with surface tension
gradients. The equilibrium contact angle $\theta_Y$ depends on the position 
of the contact point: 
\begin{equation}  \label{eqn:sigma}
\cos \theta_Y(x)=\left\{
\begin{array}{ll}
-0.8, & \quad \mbox{for}\ x<-0.8,\\
x, & \quad  \mbox{for}\ -0.8\leq x < 0.8,\\
0.8,& \quad \mbox{for}\  x\geq 0.8.
\end{array} \right.
\end{equation}
The initial configuration of the droplet is given by the rectangle 
$[-0.5, -0.25]\times[0, 0.25]$. The triangular mesh consists of $N=3036$ 
triangles and $J_\Omega=1580$ vertices. The interface contains $J_\Gamma=60$ vertices.
The time step is $\tau=2\times 10^{-4}$. Other parameters are chosen as
$\rho_1 = 1$, $\beta_1 = 0.1$, $\eta_1 = 0.1$, $\beta^*=0.1$ and $Ca=0.1$.

\begin{figure}[!t]
\centering
\includegraphics[width=1.0\textwidth]{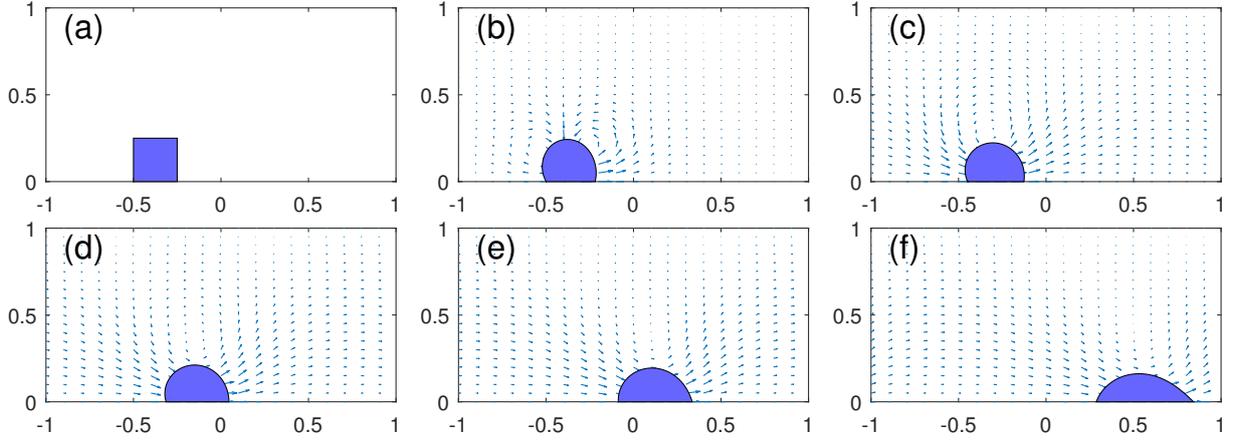}
\caption{Snapshots of the droplet migrating  on a substrate with surface 
tension gradient. (a) $t=0$; (b) $t=0.1$; (c) $t=0.4$; (d) $t=0.8$; 
(e) $t=1.3$; (f) $t=1.9$. } 
\label{Fig:Droplet}
\end{figure}

\begin{figure}[!t]
\centering
\includegraphics[width=0.8\textwidth]{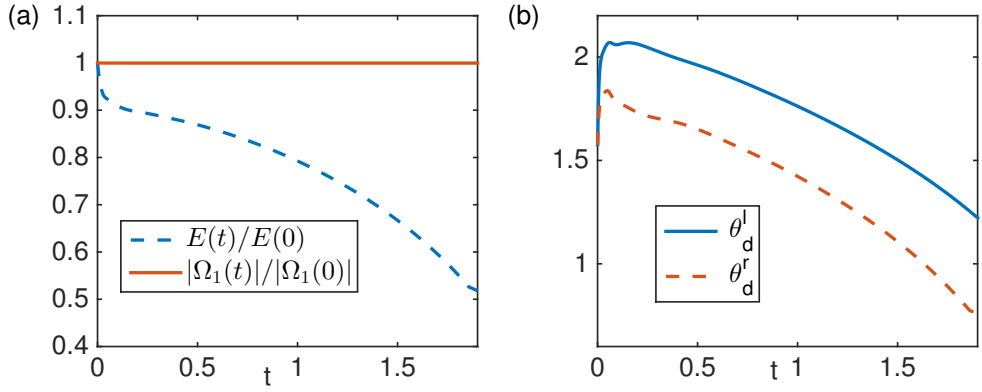}
\caption{(a) The normalized energy $E(t)/E(0)$ and the relative area 
$|\Omega1(t)|/|\Omega_1 (0)|$ versus time;
(b) The dynamic contact angles versus time.}
\label{Fig:DropletEnergy}
\end{figure}

The profiles of the droplet at several times are shown in Fig.~\ref{Fig:Droplet}. 
From the figure, we observe that the droplet first evolves into a nearly spherical 
configuration, then migrates along the substrate from the region with lower value 
of $\cos\theta_Y$ to the region with higher value of $\cos\theta_Y$ 
in order to lower the interfacial energy on the solid surface. 
The decay of the energy is shown in Fig.~\ref{Fig:DropletEnergy}.
In the figure we also show the area of the droplet and the contact angles 
versus time. We can see that the area is very well preserved.

\section{Stokes flow}\label{sec:stokes}
In this section, we consider the case in which the Reynolds number 
is small so that the flow is modeled by the time-independent 
Stokes equations in $\Omega_i\ (i=1,2)$, 
\begin{subequations}
\begin{numcases}{}
\label{eqn:stokesmodel1}
\nabla p - \nabla\cdot\tau_d =0, \\
\label{eqn:stokesmodel2}
\nabla\cdot\vec u = 0,
\end{numcases}
\end{subequations}
with the interface conditions on $\Gamma$,
\begin{subequations} \label{eq:bd1s}
\begin{align}
 \bigl[\vec u\bigr]^2_1 = 0,\\
Ca\,\bigl[\sigma\bigr]_1^2\cdot\vec n = \kappa\,\vec n,\\
\kappa = (\partial_{ss}\vec X)\cdot\vec n, \\
 \dot{\vec x}_\Gamma = \vec u|_{\vec x_\Gamma},  
\end{align}
\end{subequations}
and the same boundary and dynamic contact angle conditions as in 
\eqref{eq:bd2ab} - \eqref{eqn:bd6}, where $\sigma=p\vec I-\tau_d$.
Below we present the corresponding finite element method 
and conduct convergence tests.

\subsection{The finite element method}

The numerical method is similar to the one introduced in
section \ref{subsec:fem}. At the $m$-th time step, given a triangulation
of $\Omega_1\cup\Omega_2$, $\mathcal{T}^m$, 
which is fitted to the interface $\Gamma^m$,
we solve the following linear system for $\vec u^{m+1}\in \mathbb{U}^m$, 
$p^{m+1}\in\hat{\mathbb{P}}^m$, $\vec X^{m+1}\in V^h\times V^h_0$, 
and $\kappa^{m+1}\in V^h$,
\begin{subequations}
\begin{align}\label{eqn:sfull1}
&-\Bigl(p^{m+1},~\nabla\cdot\boldsymbol{\omega}^h\bigr)+2\,
\Bigl(\eta^m D(\vec u^{m+1}),~D(\boldsymbol{\omega}^h)\Bigr)\nonumber\\
&\qquad\qquad -\;\frac{1}{Ca}\Bigl(\kappa^{m+1}\,\vec n^m,
~\boldsymbol{\omega}^h\Bigr)_{\Gamma^m}+\; \frac{1}{l_s}
\Bigr(\beta^m\,u_s^{m+1},~\omega_s^h\Bigr)_{\Gamma_1^m\cup\Gamma_2^m}=0,
\quad\forall\boldsymbol{\omega}^h\in \mathbb{U}^m,   \\[0.5em]
\label{eqn:sfull2}
&\qquad\qquad\qquad\qquad\qquad\Bigl(\nabla\cdot\vec u^{m+1},~q^h\Bigr)=0,
\qquad\forall q^h\in \hat{\mathbb{P}}^m.\\[0.5em]
\label{eqn:sfull3}
&\qquad  \frac{1}{\tau_m} \Bigl((\vec X^{m+1}-\vec X^m)
\cdot\vec n^m,~\psi^h\Bigr)_{\Gamma^m}^h
 - \Bigl(\vec u^{m+1}\cdot\vec n^m,~\psi^h\Bigr)_{\Gamma^m}=0,
\quad\forall \psi^h\in V^h.\\[0.5em]
& \Bigl(\kappa^{m+1}\,\vec n^m,~\boldsymbol{g}^h\Bigr)_{\Gamma^m}^h
+\Bigl(\partial_s\vec X^{m+1},~\partial_s\boldsymbol{g}^h\Bigr)
_{\Gamma^m}-\cos\theta_Y\,[g_1^h(1) - g_1^h(0)] \nonumber\\
&\qquad\quad +\frac{\beta^*\,Ca}{\tau_m}\Bigl[(x^{m+1}_r- x^m_r)g_1^h(1)
+(x^{m+1}_l-x_l^m) g_1^h(0)\Bigr]=0,\quad\forall\boldsymbol{g}^h\in V^h\times V^h_0.
\label{eqn:sfull4}
\end{align}
\end{subequations}
Then  we update the triangular mesh $\mathcal{T}^m$ using the method 
introduced in section \ref{subsec:mesh} 
so that it fits to the new interface $\Gamma^{m+1}$,
and the above procedure repeats.

We can show the numerical scheme Eq.~\eqref{eqn:sfull1}-Eq.~\eqref{eqn:sfull4} 
admits a unique solution (Theorem \ref{thm:s1}) 
and satisfies a discrete energy law (Theorem \ref{thm:s2}).

\begin{thm}[Well-posedness]  \label{thm:s1}
Let $(\mathbb{U}^m,~\hat{\mathbb{P}}^m)$ satisfy the inf-sup stability condition 
\eqref{eqn:LBB} and the interface $\vec X^m(\cdot)$ satisfy the conditions in 
\eqref{eqn:assumption}. Then the numerical methods \eqref{eqn:sfull1}-\eqref{eqn:sfull4}, 
admits a unique solution.
\end{thm}

The proof is similar to the proof of Theorem \ref{th:unique1}, so is omitted.

\begin{thm}[Stability bound]  \label{thm:s2}
Let $\Bigl(\vec u^{m+1},~p^{m+1},~\vec X^{m+1},
~\kappa^{m+1}\Bigr)$ be the solution to the numerical scheme 
\eqref{eqn:sfull1}-\eqref{eqn:sfull4}. Then the following stability bound holds
\begin{eqnarray}
&&-\frac{\cos\theta_Y}{Ca}|\Gamma_1^{m+1}| +\frac{1}{Ca}|\Gamma^{m+1}| 
+ 2\tau_m\norm{\sqrt{\eta^m}D(\vec u^{m+1})}_0^2  + 
\frac{\tau_m}{l_s}\Bigl(\beta^m\,u_s^{m+1},~u_s^{m+1}\Bigr)
_{\Gamma^m_1\cup\Gamma_2^m} \nonumber\\ 
&&\qquad\qquad +\frac{\beta^*}{\tau_m}\Bigl[(x^{m+1}_r-x^m_r)^2
+(x^{m+1}_l-x^m_l)^2\Bigr]\leq -\frac{\cos\theta_Y}{Ca}|\Gamma_1^{m}| 
+ \frac{1}{Ca}|\Gamma^{m}|.
\label{eqn:senergybounds}
\end{eqnarray}
Moreover, for $k \ge 1$, we have 
\begin{eqnarray}
&&-\frac{\cos\theta_Y}{Ca}|\Gamma_1^{k}| + \frac{1}{Ca}|\Gamma^{k}| 
+ \sum_{m=0}^{k-1}2\tau_m\norm{\sqrt{\eta^m}D(\vec u^{m+1})}_0^2 
+\sum_{m=0}^{k-1}\frac{\tau_m}{l_s}\Bigl(\beta^m\,u_s^{m+1},~u_s^{m+1}\Bigr)
_{\Gamma^m_1\cup\Gamma_2^m}\nonumber\\
&&\qquad\qquad +\sum_{m=0}^{k-1}\frac{\beta^*}{\tau_m}
\Bigl[(x^{m+1}_r-x^m_r)^2+(x^{m+1}_l-x^m_l)^2\Bigr]\leq 
-\frac{\cos\theta_Y}{Ca}|\Gamma_1^{0}| + \frac{1}{Ca}|\Gamma^{0}|.
\label{eqn:senergybounds2}
\end{eqnarray}

\begin{proof}
Choosing $\boldsymbol{\omega}^h=\vec u^{m+1}$, $q^h = p^{m+1}$, 
$\psi^h = \frac{1}{Ca}\kappa^{m+1}$ and 
$\boldsymbol{g}^h = \frac{1}{Ca}(\vec X^{m+1}-\vec X^m)$ 
in \eqref{eqn:sfull1}-\eqref{eqn:sfull4}, then combining the equations yields 
\begin{eqnarray}
&& 2\Bigl(\eta^m D(\vec u^{m+1}),~D(\vec u^{m+1})\Bigr)
+ \frac{1}{l_s}\left(\beta^m\,u_s^{m+1},~u_s^{m+1}\right)_{\Gamma_1^m\cup\Gamma_2^m}
 +\frac{1}{Ca\cdot \tau_m}\Bigl(\partial_s\vec X^{m+1},~\partial_s
(\vec X^{m+1}-\vec X^m)\Bigr)_{\Gamma^m}    \nonumber\\
&& -\frac{\cos\theta_Y}{Ca\cdot\tau_m}
\Bigl[(x_r^{m+1}-x_l^{m+1})-(x_r^m-x_l^m)\Bigr] 
+\frac{\beta^*}{ (\tau_m)^2}
\Bigl[(x_r^{m+1}-x_r^m)^2+(x_l^{m+1}-x_l^m)^2\Bigr]=0.
\label{eqn:senergybound1}
\end{eqnarray}
Eq.~\eqref{eqn:senergybounds} immediately follows by noting 
Eq.~\eqref{eqn:energybound4}. 
By summing up Eq.~\eqref{eqn:senergybounds} for $m$ from $0$ to $k-1$, 
we obtain the energy dissipation law Eq.~\eqref{eqn:senergybounds2}.
\end{proof}
\end{thm}

In contrast to the numerical scheme in \eqref{eqn:full1}-\eqref{eqn:full4} 
for the Navier-Stokes equations, the interpolation step of the velocity 
and density fields from $\mathcal{T}^m$ to the new mesh $\mathcal{T}^{m+1}$ 
is not needed for Stokes flow. This allowed us to prove the global 
energy dissipation law in \eqref{eqn:senergybounds2}. 
Similar work for the two-phase Stokes flow without contact lines 
has been done in Ref. \cite{Agnese16}; 
there the method was shown to be unconditionally stable.

\begin{table}[t]
\centering
\def\temptablewidth{0.95\textwidth}
\vspace{-12pt}
\caption{Error of the numerical solution and the rate of convergence for the fluid 
interface modeled using the Stokes equations. $h=1/J_\Gamma$ and $\tau$ are the 
mesh size and the time step, respectively, where $h_0=1/36$ and $\tau_0=0.01$.
The numerical results are obtained using the {\rm P2-P0} elements (upper panel)
 and the {\rm P2-(P1+P0)} elements (lower panel).}
{\rule{\temptablewidth}{1pt}}
\begin{tabular*}{\temptablewidth}{@{\extracolsep{\fill}}ccccccc}
$(h,\ \tau)$ & $e_{h,\tau}(t=0.2) $ & order &$e_{h,\tau}(t=1.0)$ 
& order &$e_{h,\tau}(t=4.0) $ & order  \\ \hline
$(h_0, \tau_0)$ & 4.10E-3 & - &4.20E-3 &-& 4.19E-3 &- \\ \hline
$(\frac{h_0}{2}, \frac{\tau_0}{2^2})$ & 1.15E-3 & 1.83 &1.20E-3 &1.81& 1.20E-3 &1.80 
\\ \hline
$(\frac{h_0}{2^2}, \frac{\tau_0}{2^4})$ & 3.08E-4 & 1.90 &3.23E-4 &1.89& 3.22E-4 &1.90 
 \end{tabular*}
{\rule{\temptablewidth}{1pt}}
{\rule{\temptablewidth}{1pt}}
\begin{tabular*}{\temptablewidth}{@{\extracolsep{\fill}}ccccccc}
$(h,\ \tau)$ &$e_{h,\tau}(t=0.2) $ & order &$e_{h,\tau}(t=1.0)$ 
& order &$e_{h,\tau}(t=4.0) $ & order  \\ \hline
$(h_0, \tau_0)$ & 4.13E-3 & - &4.15E-3 &- &4.13E-3& - \\ \hline
$(\frac{h_0}{2},\frac{\tau_0}{2^2})$& 1.18E-3 & 1.81 &1.18E-3 &1.81& 1.18E-3 &1.81 
\\ \hline
$(\frac{h_0}{2^2}, \frac{\tau_0}{2^4})$ & 3.13E-4 & 1.91 &3.16E-4 &1.90 & 3.15E-4 &1.91
 \end{tabular*}
{\rule{\temptablewidth}{1pt}}
\label{tb:order0}
\end{table}

\subsection{Convergence test}

We investigate the accuracy and the convergence rate of the numerical method 
using the same example in section \ref{subsec:con}, with the parameters
$\beta_1=0.1$, $\eta_1=10$, $\beta^*=0.1$, $\theta_Y=2\pi/3$, $l_s = 0.1$
and $Ca=0.01$. The numerical results are summarized in Table.~\ref{tb:order0},
where the errors of the fluid interface are computed using Eq. \eqref{errorn}.
We can clearly observe the convergence for both P2-P0 and P2-(P1+P0) elements. 
The convergence rates approach 2 as the mesh is refined.
 

\section{Conclusions}
\label{sec:con}
In this work, we have developed an efficient energy-stable numerical method 
for two-phase fluids with moving contact lines. The method combines
the finite element method for the Navier-Stokes/Stokes equations with 
a semi-implicit parametric finite element method for the dynamics of the fluid interface.
 We used the moving mesh approach such that the evolving fluid interface 
remains fitted to the triangular mesh. At each time step, the new mesh is constructed 
based on the mesh at the previous time step by solving an elastic equation 
with proper boundary conditions for the displacements 
of the internal nodes.   

The contact line condition in the model relates the dynamic contact angle of the interface
to the contact line velocity. It is a non-trivial task to properly impose this condition 
in numerical simulations. In this work, we formulated it as a time-dependent Robin-type of 
boundary condition for the fluid interface so it is naturally imposed 
in the weak form of the governing equations.

For the Navier-Stokes equations, we showed that the numerical scheme 
obeys a similar energy law as the continuum model but up to an error due to 
the interpolation of the numerical solutions on the moving mesh.
For Stokes flow, the interpolation is not needed so we were able to prove the global
unconditional stability in terms of the energy.
Numerical simulations have demonstrated the convergence and accuracy of the numerical methods.
For Stokes flows, the convergence rate for the interface dynamics reaches about 2 
as the mesh is refined. However, for Navier-Stokes equations, the numerical solution
is polluted by the interpolation error, and the order of convergence is unstable.

The current work focused on systems in two dimensions. In the future, we intend to 
extend the numerical method to systems in three dimensions and also more challenging 
problems such electro-wetting, contact line dynamics on elastic substrate, etc.

\section*{Acknowledgement} 
The work was partially supported by Singapore MOE AcRF grants (R-146-000-267-114,
R-146-000-285-114) and NSFC (NO. 11871365).

\section*{Reference}
\bibliographystyle{elsarticle-num}
\bibliography{thebib}
\end{document}